\documentclass[11pt,a4paper]{article}
\usepackage{jheppub}
\usepackage[T1]{fontenc}
\usepackage{amssymb}
\usepackage{mathtools}
\usepackage{stackengine}
\usepackage{scalerel}

\usepackage{hyperref}
\usepackage{epsfig}
\usepackage{amsmath,latexsym,amssymb}
\usepackage{graphicx}
\usepackage{braket}
\usepackage{pdfsync}

\usepackage{framed}
\usepackage{mathtools}

\usepackage{color}
\usepackage{pstricks}

\usepackage{jheppub}
\usepackage[T1]{fontenc}
\usepackage{amssymb}
\usepackage{mathtools}
\usepackage{amsmath}
\usepackage{bm}
\usepackage{stackengine}

\usepackage{scalerel}
\usepackage{eufrak}
\usepackage{framed}

\usepackage{mathtools}

\usepackage{amsmath}
\usepackage{bbold}
\usepackage{bm}
\usepackage{latexsym}
\usepackage{braket}
\usepackage{slashed}
\usepackage{graphicx,booktabs,multirow}
\usepackage{eufrak}
\numberwithin{equation}{section}

\usepackage{tikz}

\usetikzlibrary{
    calc,
    arrows.meta, 
    patterns,      
    decorations.pathreplacing,  
    shapes.misc 
}
\usepackage{ulem}

\makeatletter
\newcommand{\doublewidetilde}[1]{{%
  \mathpalette\double@widetilde{#1}%
}}
\newcommand{\double@widetilde}[2]{%
  \sbox\z@{$\m@th#1\widetilde{#2}$}%
  \ht\z@=.9\ht\z@
  \widetilde{\box\z@}%
}
\makeatother

\usepackage{hyperref}
\usepackage{slashed}
\usepackage{epsfig}
\usepackage{amsmath,latexsym,amssymb}
\usepackage{graphicx}
\usepackage[latin1]{inputenc}
\usepackage{braket}
\usepackage{pdfsync}

\usepackage{framed}

\usepackage{mathtools}

\def\be{\begin{equation}}
\def\ee{\end{equation}}
\def\ba{\begin{eqnarray}}
\def\ea{\end{eqnarray}}

\def\k{\kappa}

\def\p{\pi}

\newcommand{\comment}[1]{}

\def\j{\bm{j}}
\def\jj{\bm{j'}}
\def\p{\partial}
\newcommand{\eea}{\end{eqnarray}}

\author{
Audrey Lindsay${}^{1}$ ,\, Tomasz R.\ Taylor${}^{1,2}$\\[0.5cm]
 $^1${\it Department of Physics,
  Northeastern University, Boston, MA 02115, USA}\\
  $^2${\it Faculty of Physics, University of Warsaw, Pasteura 5, 02-093 Warsaw, Poland}\\[0.2cm]
}

\emailAdd{a.lindsay@northeastern.edu}
\emailAdd{taylor@neu.edu}

\title{\boldmath {Symmetries of de Sitter Particles and Amplitudes} \unboldmath}

\abstract{We discuss the symmetry aspects of quantum field theory in global four-dimensional de Sitter spacetime linked to $SO(1,4)$ isometries. For the unitary irreducible representations relevant to elementary particles, we obtain explicit transformation laws for the symmetry generators acting on one-particle states in a basis adapted to the $SU(2) \times SU(2)'$ decomposition of the Hilbert space. Using these results, we derive the corresponding Ward identities and demonstrate how global spacetime symmetries constrain de Sitter scattering amplitudes. We show that the Poincar\'e algebra and flat-space Ward identities are recovered in the large-momentum limit.}

\gdef\@fpheader{}
\makeatother

\begin{document}
\maketitle
\section{Introduction}
The S-matrix encodes the amplitudes describing  the scattering processes of physical particles in ambient spacetime. Its structure reflects the  symmetries of spacetime. In Minkowski space,  Poincar\'e symmetry allows one to construct the scattering amplitudes in a way that does not depend on the choice of an observer's reference frame. This symmetry becomes most transparent when the S-matrix is written in the momentum basis: the energy-momentum is conserved and the amplitudes depend on four-momenta through Lorentz-invariant kinematic variables. In some other bases, for example in the celestial (boost) basis, Poincar\'e symmetry is not so explicit, but it can always be displayed using Ward identities. There is one example of curved spacetime -- the maximally symmetric (global) de Sitter spacetime -- where  the observer's choice can be handled in the same way as in the flat case. De Sitter spacetime has the topology of $\mathbb{R}\times S^3$. By using $SO(1,4)$ de Sitter symmetry, any inertial observer, at any arbitrary moment of his/her proper time, can be placed at the North pole of $S^3$.  The most natural basis that he or she can use to describe the scattering amplitudes is the basis of unitary $SO(1,4)$ representations.  In this work, we discuss the Ward identities that relate the scattering amplitudes expressed in this basis. Although our discussion was prompted by the recent construction of the scattering amplitudes in \cite{Taylor:2024vdc,Taylor:2025deepIR}, such Ward identities are completely universal because the only assumption used in the derivation is the existence of a de Sitter invariant S-matrix.\footnote{For earlier work on the de Sitter invariant S-matrix, see \cite{Marolf:2012kh,Parikh:2002py}.}

The paper is organized as follows. In section 2 we review the geometry of four-dimensional de Sitter spacetime and construct the scalar mode functions, which are linear combinations of hyperspherical harmonics on $S^3$ weighted by time-dependent Ferrers functions. Section 3 summarizes the $so(1,4)$ isometry algebra and presents the structure of the global Killing generators. In section 4 we describe the unitary irreducible representations (UIRs) relevant for scalar fields, massive and massless fermions, gauge bosons, and gravitons, and obtain the explicit action of symmetry generators on one-particle states. Section 5 derives the corresponding Ward identities and illustrates the constraints they impose on scattering processes through several simple examples. We consider an observer at the North pole and show that in the flat limit, the de Sitter generators reduce to the usual Poincar\'{e} generators acting on plane-wave states, thereby recovering the expected Minkowski Ward identities. Section 6 contains our conclusions. In Appendix A, we state the explicit form of our Killing vectors; in Appendix B we provide definitions of the special functions used throughout and the relevant identities used in intermediate steps, particularly those entering the computation of generator actions on one-particle states in the scalar principal series representation.

\section{De Sitter geometry, wavefunctions and Hilbert space}
In 1961, Jacques Dixmier classified all unitary irreducible representations of the $SO(1,4)$ de Sitter symmetry group.\footnote{For more recent work, see \cite{rep2,rep21,rep3,rep4}.}
He showed that the Hilbert space can be decomposed as the sum $\oplus_{(j, j')\in \Gamma}{\cal H}_{j,j'}$ of  $SU(2)\times SU(2)'$ representations ${\cal H}_{j,j'}$,  $(j,j')\in \Gamma$, with the set $\Gamma$ depending on the  $SO(1,4)$ representation content. In quantum field theory, the states of Hilbert space represent quantum particles. Only a few representations, however, are physically relevant; we will describe them later. Their physical properties are described by wavefunctions constructed by solving classical field equations.  In order to construct the S-matrix directly in a given basis, it is convenient to use the wavefunctions expressed in the coordinates in which the Lie derivatives along the Killing vectors make the action of the symmetry generators on the basis vectors as simple as possible.

A four-dimensional de Sitter space $dS_4$ is realized as the embedding of a one-sheeted hyperboloid (topologically $\mathbb{R}\times S^3$) in $D=5$ Minkowski space, described by 
\begin{equation}
-X_0^2 + X_1^2 + X_2^2 + X_3^2 + X_4^2 = \ell^2,
\end{equation}
where $\ell>0$ is the de Sitter radius. From here, we set $\ell = 1$.  We use conformal time coordinate  $t \in [-\pi/2, \pi/2]$
and the Hopf fibration of $S^3$ to parametrize it by three Hopf angles, a.k.a.\ toroidal coordinates,  $\chi \in [0, \pi/2]$ and $\theta, \varphi \in [0, 2\pi]$ \cite{Lehoucq:2002E}. As explained at the end of this  section, these coordinates are suitable for Dixmier's $SU(2)\times SU(2)'$ ``isospin'' decomposition of the Hilbert space. They are related to the embedding space coordinates as follows:
\begin{equation}
\begin{aligned}
X^0 &= \tan t,\\
X^1 &= \sec t\,\cos\chi\,\cos\theta,\\
X^2 &= \sec t\,\cos\chi\,\sin\theta,\\
X^3 &= \sec t\,\sin\chi\,\cos\varphi,\\
X^4 &= \sec t\,\sin\chi\,\sin\varphi.
\end{aligned}
\label{eq:coordinates}
\end{equation}
In global conformal coordinates, the $dS_4$ metric is
\begin{equation}
ds^2 = \frac{1}{\cos^2{t}}(-dt^2 + d\Omega^2),
\end{equation}
where
\begin{equation}
d\Omega^2 = d\chi^2 + \cos^2{\chi} d\theta^2 + \sin^2{\chi}d\varphi^2 \label{s3}
\ee
is the metric on a unit $S^3$.

The integer spin wavefunctions can be constructed from the solutions of  Klein-Gordon equation.
{}For a scalar field $\phi(t, \Omega)$ with Lagrangian mass $m$, it reads
\begin{equation}
\Big(\partial_t^2 + 2\tan{t} \partial_t - \Delta_{S^3} + \frac{m^2}{\cos^2{t}}\Big) \phi(t, \Omega) = 0,
\end{equation}
where $\Delta_{S^3}$ is the Laplacian on $S^3$.  The eigenfunctions of $ \Delta_{S^3} $ are known as hyperspherical harmonics $ Y_{klm}(\Omega) $ and satisfy
\begin{equation}
\Delta_{S^3} Y_{kl m}(\Omega) = -k(k+2)Y_{kl m}(\Omega). 
\end{equation}
In  toroidal coordinates $(\chi, \theta, \varphi)$, they are given by \cite{Lehoucq:2002E}
\begin{equation}
Y_{{klm}}(\Omega) = N_{klm} \cos^L{\chi}\sin^M{\chi}P_{d}^{(M, L)}(\cos{2\chi})e^{il \theta}e^{i m \varphi},\label{harms}
\end{equation}
where $P_{d}^{(M,L)}$ are the Jacobi polynomials defined in Appendix B.2. The integers $k\ge 0,~l,$ and $m$ determine the parameters
\be
L = |l|, ~M = |m|, ~d = \frac{1}{2}(k-M-L).\label{set1}\ee
For each $k$, $l$ and $m$ are constrained by
 \be \Gamma_k:~~~ M + L \leq k, ~M+L = k \mod{2}, \label{set}\ee
so that $d\ge 0$ is also an integer. These functions are normalized with respect to the $S^3$ metric (\ref{s3}) with
\begin{equation}
N_{klm} = \frac{i^M}{i^m}\frac{1}{2\pi}\sqrt{\frac{2\,(k+1)\,d!\,(L+M+d)!}{(L+d)!\,(M+d)!}}.\label{normh}
\end{equation}

{}For a single harmonic mode
\be \phi_{klm}(t,\Omega)=u_k(t)Y_{{klm}}(\Omega) \ee
with the time-dependent envelope $u_k(t),$
the Klein-Gordon equation reduces to 
\begin{equation}
u''_k(t) + 2\tan{t}\,u'_k(t) + \Big[k(k+2) + \frac{m^2}{\cos^2{t}}\Big]u_k(t) = 0.
\end{equation}
The solutions can be classified as positive or negative frequency modes, according to the criteria described in \cite{Taylor:2024vdc,Taylor:2025deepIR}. The properly normalized positive frequency modes are given by
\begin{align}
\phi_{klm}(t,\Omega)^{(+)} = \frac{e^{- \frac{\mu\pi}{2}} e^{-i\frac{\pi}{4}} \sqrt{\pi}}{2}&
\sqrt{\Gamma(k+\frac{3}{2}+i\mu)\over \Gamma(k+\frac{3}{2}-i\mu)}(-i)^{k}
\nonumber \\ &
\times \sqrt{\cos t}^3 \big(P^{-i\mu}_{k+\frac{1}{2}}(\sin t) -\frac{2i}{\pi} Q^{-i\mu}_{k+\frac{1}{2}}(\sin t) \big) Y_{klm}(\Omega)\, ,\label{lwap}
\end{align}
where $P$ and $Q$ are Ferrers functions defined in Appendix B.1 and
\be \mu=\sqrt{m^2-\frac{9}{4}}\ .\label{mkk}\ee
The negative frequency modes are obtained by complex conjugation:
\be
\phi_{klm}(t,\Omega)^{(-)}=\phi_{klm}(t,\Omega)^{(+)*}
.\label{lwam}
\ee
These wavefunctions are normalized with respect to the Klein-Gordon scalar product
\be
(f|g)_{\text{\scriptsize KG}} =i\int (f^*\partial_t g-g\, \partial_t f^*)\, d\Omega\label{kgs}\ee
in the following way:
\begin{equation}
(\phi^{(\pm)}_{kl m}| \phi^{(\pm)}_{k' l' m'})_{\text{\scriptsize KG}} = \pm\delta_{kk'}\delta_{l l'}\delta_{m m'} \qquad (\phi^{(\pm)}_{kl m}| \phi^{(\mp)}_{k' l' m'})_{\text{\scriptsize KG}} = 0\ .
\end{equation}
As explained in \cite{Taylor:2024vdc,Taylor:2025deepIR}, for real $\mu$, the wavefunctions (\ref{lwap}) describe scalar particles belonging to the principal series representation of $SO(1,4)$.  The complementary series can be reached by extrapolating to imaginary $\mu$. Higher spin wavefunctions can be constructed in a similar way.

At this point, we can make a connection to Dixmier's $SU(2) \times SU(2)'$ isospin decomposition of the Hilbert space, ${\cal H}= \oplus_{(j, j')\in \Gamma}{\cal H}_{j,j'}$. The basis of ${\cal H}_{j,j'}$ consists of 
\be  
\big|(\j\,\nu)(\,\jj\nu')\big\rangle
\ ,\quad\nu=-\j,-\j+1,\dots,\j ,~~\nu'=-\jj,-\jj+1,\dots,\jj\ ,\label{dst}\ee
with  $\j\ge 0$ and $\jj\ge0$ taking half-integer values. If we specify to the principal series, then $\j=\jj$, and the bases occupy the straight line shown in Figure 1, starting at $(\j,\jj)=(0,0)$, with $\j=\jj$ increasing in half-integer increments. In section 4, we will show that upon identification
\be k=\j+\jj =2\j=2\jj\ , ~~~~l=\nu+\nu'\ , ~~~~m=\nu-\nu'\ ,\label{jkrel}\ee
\be  \big|k\;l\,m\big\rangle\, \label{bas2}\equiv\,
\big|(\j\,\nu)(\,\jj\nu')\big\rangle
 ,\ee
the wavefunctions $\phi_{klm}(t,\Omega)^{(+)}$, as written in (\ref{lwap}), describe the basis states  (\ref{bas2}).\footnote{In Dixmier's notation, $ f^{j,j'}_{\nu,\nu'}\equiv
\big|(\j\,\nu)(\,\jj\nu')\big\rangle$.} There is a similar relation for higher spin states.
\section{De Sitter isometry algebra}
De Sitter isometries are generated by ten Killing vectors $K_{AB}=-K_{BA}$, $A,B=0,\dots,4$, whose explicit forms in toroidal coordinates are given in Appendix A.
The Killing equations imply that for any two solutions $\phi_{1,2}$ of the Klein-Gordon equation, 
\be (\phi_1|K_{AB}\phi_2)_{\text{\scriptsize KG}}+(K_{AB}\phi_1|\phi_2)_{\text{\scriptsize KG}} =0 ~~~\forall\, A,B,\ee
where
\be K_{AB}\phi\equiv {\cal L}_{K_{AB}}\phi=d\phi(K_{AB}).\ \ee
Therefore, $K_{AB}$ are antihermitean with respect to the Klein-Gordon product and generate unitary transformations.
They satisfy the  $so(1,4)$ algebra 
\begin{equation}
[K_{AB}, K_{CD}] = \eta_{BC}K_{AD}-\eta_{AC}K_{BD}-\eta_{BD}K_{AC}+
\eta_{AD}K_{BC},\label{kcom}
\end{equation}
with $\eta =\makebox{diag}(-++++)$. Following Dixmier, we introduce the operators
\begin{align}
   & L = -K_{12}-K_{34} = -\partial_{\theta} - \partial_{\varphi},\nonumber \\
   & L'  = -K_{12}+K_{34} = -\partial_{\theta} + \partial_{\varphi},\nonumber \\
   & X_{\pm \alpha }  = (-K_{13}+K_{24})\pm i(-K_{14}-K_{23}) = e^{\pm i \theta}e^{\pm i \varphi}(- \partial_{\chi} \pm i\tan{\chi}\,\partial_{\theta} \mp i\cot{\chi}\,\partial_{\varphi}), \nonumber \\
    &X_{\pm \beta}  = (-K_{13}-K_{24})\pm i(-K_{14}+K_{23})= e^{\mp i \theta}e^{\pm i \varphi}(- \partial_{\chi} \mp i\tan{\chi}\,\partial_{\theta} \mp i\cot{\chi}\,\partial_{\varphi}),\nonumber \\
    &X_{\pm \gamma}  = K_{01} \pm i K_{02}  = 
    e^{\pm i \theta}(\cos{t}\cos{\chi}\,\partial_t - \sin{t}\sin{\chi}\, \partial_{\chi} \pm i \sin{t}\sec{\chi}\,\partial_{\theta}),\nonumber \\
    &X_{\pm \delta}  =  K_{03}\pm iK_{04} = e^{\pm i\varphi} (\cos{t}\sin{\chi}\,\partial_t + \sin{t}\cos{\chi}\,\partial_{\chi} \pm i \sin{t}\csc{\chi}\,\partial_{\varphi}).\label{opdef}
\end{align}
The commutation relations are written below:
\begin{equation}
\begin{cases}
[L, L'] = 0\\
[L, X_{\pm \alpha}] = \mp 2 i X_{\pm \alpha} \\
[L, X_{\pm \beta}] = 0\\
[L, X_{\pm \gamma}] = \mp i X_{\pm \gamma}\\
[L, X_{\pm \delta}] = \mp X_{\pm\delta}
\end{cases}\label{xx1}
\end{equation}

\begin{equation}
\begin{cases}
[L', X_{\pm \alpha}] = 0 \\
[L', X_{\pm \beta}] = \pm 2 i X_{\pm \beta}\\
[L', X_{\pm \gamma}] = \mp i X_{\pm \gamma}\\
[L', X_{\pm \delta}] = \pm i X_{\pm\delta}
\end{cases}\label{xx2}
\end{equation}

\begin{equation}
\begin{cases}
[X_{\alpha}, X_{- \alpha}] = -4iL \\
[X_{\alpha}, X_{\pm \beta}] = 0 & [X_{-\alpha}, X_{\pm \beta}] = 0\\
[X_{\alpha}, X_{\gamma}] = 0 & [X_{-\alpha}, X_{\gamma}] = 2X_{-\delta}\\
[X_{\alpha}, X_{-\gamma}] = 2X_{\delta} & [X_{-\alpha}, X_{-\gamma}] = 0\\
[X_{\alpha}, X_{\delta}] = 0 & [X_{-\alpha}, X_{\delta}] = -2X_{-\gamma}\\
[X_{\alpha}, X_{-\delta}] = -2X_{\gamma} & [X_{-\alpha}, X_{-\delta}] = 0
\end{cases}\label{xx3}
\end{equation}

\begin{equation}
\begin{cases}
[X_{\beta}, X_{- \beta}] = 4iL'\\
[X_{\beta}, X_{\gamma}]  = 2X_{\delta} & [X_{-\beta}, X_{\gamma}] = 0\\
[X_{\beta}, X_{-\gamma}] = 0 & [X_{-\beta}, X_{-\gamma}] = 2X_{-\delta} \\
[X_{\beta}, X_{\delta}] =  0 & [X_{-\beta}, X_{\delta}] = -2X_{\gamma}\\
[X_{\beta}, X_{-\delta}] = -2X_{-\gamma} & [X_{-\beta}, X_{-\delta}] = 0
\end{cases}\label{xx4}
\end{equation}

\begin{equation}
\begin{cases}
[X_{\gamma}, X_{- \gamma}] = i(L+L')  \\
[X_{\delta}, X_{- \delta}] = i(L-L')\\
[X_{\gamma}, X_{\delta}] = -X_{\alpha}  & [X_{-\gamma}, X_{\delta}] = -X_{\beta} \\
[X_{\gamma}, X_{-\delta}] = -X_{-\beta} & [X_{-\gamma}, X_{-\delta}] =- X_{-\alpha} \\
\end{cases}\label{xx5}
\end{equation}
The operators ($L, X_{\pm\alpha}$) and ($L', X_{\pm\beta}$) generate the $su(2)$ and $su(2)'$ subalgebras, respectively; see (\ref{xx1}) and (\ref{xx2}). They transform states within the multiplets labeled by isospin $(\j,\jj)$. On the other hand, as shown by Dixmier, and as elaborated in the next section, $X_{\pm\gamma}$ and  $X_{\pm \delta}$ raise or  lower $\j$ and  $\j'$ by one half. For the principal series, this corresponds to $k\to k\pm 1$.

\section{UIRs of elementary particles}
\subsection{Spin 0 principal series}
The first step towards deriving Ward identities is to determine how the symmetry operators act on one-particle states. For any representation, one starts from the mode decomposition of the free quantum field. Here, we specify to the scalar field
\begin{equation}
    \phi(t, \Omega) = \sum_{k=0}^\infty\,\sum_{l,m\in \Gamma_k}\big[a_{kl m} \phi^{(+)}_{kl m}(t, \Omega)+ a^{\dagger}_{klm}\phi^{(-)}_{klm}(t, \Omega)\big]\ ,\label{sfield}
\end{equation}
with the set $\Gamma_k$ defined in (\ref{set1})-(\ref{set}) and the wavefunctions in (\ref{lwap})-(\ref{lwam}). Here $a^\dagger_{kl m}$ and  $a_{kl m}$ are the creation and annihilation operators, respectively, satisfying the canonical commutation relations:
\be [a_{kl m}, a^\dagger_{k'l' m'}]=\delta_{kk'}\delta_{l l'}\delta_{m m'} \ ,\qquad  [a_{kl m},a_{k'l 'm'}]=0\ ,  \qquad[a^\dagger_{kl m} , a^\dagger_{k'l 'm'}]=0\ .\label{acom}\ee
The de Sitter invariant (Bunch-Davies \cite{Bunch:1978yq}) vacuum is defined by
\be a_{kl m} \big|0\big\rangle=0~~~ \forall\, k,l,m,\ee
and one-particle states by
\be
 \big|k\;l\,m\big\rangle=a^\dagger_{kl m} \big|0\big\rangle\ .\label{stdef}\ee
Their orthonormality follows from the normalization $\langle 0|0\rangle=1$ and the commutation relations (\ref{acom}):
\be
 \big\langle k\;l\,m\,\big|k'\;l'\,m'\big\rangle=\delta_{kk'}\delta_{l l'}\delta_{m m'} \ \label{orthr}.\ee

Our goal is to determine how the symmetry generators act on one-particle states (\ref{stdef}). To that end, we compute the directional (Lie) derivatives of the wavefunctions (\ref{lwap}) along the vector fields (\ref{opdef}). As a result of tedious computations, using the properties of Jacobi polynomials and Ferrers functions listed in Appendix B, we obtain: 
\begin{equation}
\begin{aligned}
&d\phi^{(+)}_{klm}(L) = -i(l+m)\phi^{(+)}_{kl m}\ ,\\[1mm]
&d\phi^{(+)}_{klm}(L') = -i(l-m)\phi^{(+)}_{kl m}\ ,\\[1mm]
&d\phi^{(+)}_{klm}(X_{\pm \alpha})= A_{\pm \alpha}(k, l, m)\phi^{(+)}_{k, l \pm 1, m\pm 1}\ ,\\[1mm]
&d\phi^{(+)}_{klm}(X_{\pm \beta})= A_{\pm \beta}(k, l, m)\phi^{(+)}_{k, l \mp 1, m\pm 1}\ ,\\[1mm]
&d\phi^{(+)}_{klm}(X_{\pm \gamma}) = A_{\pm \gamma}(k, l, m)\phi^{(+)}_{k+1, l \pm 1, m} + D_{\pm \gamma}(k, l, m)\phi^{(+)}_{k-1, l \pm 1, m}\ ,\\[1mm]
&d\phi^{(+)}_{klm}(X_{\pm \delta})= A_{\pm \delta}(k, l, m)\phi^{(+)}_{k+1, l , m\pm 1} + D_{\pm \delta}(k, l, m)\phi^{(+)}_{k-1, l, m\pm 1}\ ,\\[1mm]
\end{aligned}
\label{phider}
\end{equation}
with the coefficients\\[1mm]
\begin{equation}
\begin{aligned}
A_{\pm \alpha}(k, l, m) & = \pm \sqrt{(k \mp l \mp m)(k\pm l \pm m + 2)},\\[3mm]
A_{\pm \beta}(k, l, m) & = \pm \sqrt{(k\pm l \mp m)(k\mp l \pm m +2)},\\[2mm]
A_{\pm \gamma}(k, l, m) & = - i \sqrt{\frac{(k \pm l \pm m + 2)(k \pm l \mp m + 2)(\mu^2 + (k + \frac{3}{2})^2)}{4(k+2)(k+1)}},\\
D_{\pm \gamma}(k, l, m) & = -i\sqrt{\frac{(k \mp l \mp m)(k \mp l \pm m)(\mu^2 + (k + \frac{1}{2})^2)}{4k(k+1)}},\\
A_{\pm \delta}(k, l, m) & = \mp i\sqrt{\frac{(k \pm l \pm m + 2)(k \mp l \pm m + 2)(\mu^2 + (k+\frac{3}{2})^2)}{4(k+2)(k+1)}},\\
D_{\pm \delta}(k, l, m) & = \pm i\sqrt{\frac{(k\mp l \mp m)(k \pm l \mp m)(\mu^2+(k+\frac{1}{2})^2)}{4k(k+1)}}.\\
\end{aligned}\label{ccoef}
\end{equation}
Upon complex conjugation,
\begin{equation}
\begin{aligned}
&A^{\dagger}_{\pm \alpha}(k, l, m)=  A_{\pm \alpha}(k, l, m)=-A_{\mp \alpha}(k, l \pm 1, m\pm 1)\ , \\
 &A^{\dagger}_{\pm \beta}(k, l, m)= A_{\pm \beta}(k, l, m)=-A_{\mp \beta}(k, l \mp 1, m \pm 1)\ ,  \\
&  A_{\pm \gamma}^{\dagger}(k, l, m)= -A_{\pm \gamma}(k, l, m)=-D_{\mp \gamma}(k+1, l \pm 1, m)\ ,\\
& D^{\dagger}_{\pm \gamma}(k, l, m)=-D_{\pm \gamma}(k, l, m)=-A_{\mp \gamma}(k-1, l \pm 1, m)\ , \\
&A_{\pm \delta}^{\dagger}(k, l, m)= -A_{\pm \delta}(k, l, m)=-D_{\mp \delta}(k+1, l, m\pm1)\ , \\
&D_{\pm \delta}^{\dagger}(k, l, m)=-D_{\pm \delta}(k, l, m)=-A_{\mp \delta}(k-1, l, m \pm 1)\ .
\end{aligned}\label{cconj}
\end{equation}

At the quantum level, with the symmetry generators represented by the charge operators $Q$, the following commutators generate the transformations (\ref{phider}) of the quantum fields (\ref{sfield}):
\begin{equation}
\begin{aligned}
&[Q_L, a^{\dagger}_{kl m}]  = -i(l+m)a^{\dagger}_{kl m}\ ,\\
&[Q_{L'}, a^{\dagger}_{kl m}]  = -i(l-m)a^{\dagger}_{kl m}\ ,\\
&[Q_{\pm \alpha}, a^{\dagger}_{kl m}]  = A_{\pm \alpha}(k, l, m)a^{\dagger}_{k, l \pm 1, m\pm 1}\ ,\\
&[Q_{\pm \beta}, a^{\dagger}_{kl m}]  = A_{\pm \beta}(k, l, m)a^{\dagger}_{k, l \mp 1, m\pm 1}\ ,\\
&[Q_{\pm \gamma}, a^{\dagger}_{kl m}]  = A_{\pm \gamma}(k, l, m)a^{\dagger}_{k+1, l \pm 1, m} + D_{\pm \gamma}(k, l, m)a^{\dagger}_{k-1, l \pm 1, m}\ ,\\
&[Q_{\pm \delta}, a^{\dagger}_{kl m}]  = A_{\pm \delta}(k, l, m)a^{\dagger}_{k+1, l , m\pm 1} + D_{\pm \delta}(k, l, m)a^{\dagger}_{k-1, l, m\pm 1}\ .
\end{aligned}\label{qac}
\end{equation}
Similarly,
\begin{equation}
\begin{aligned}
&[Q_L, a_{kl m}]  = i(l+m)a_{kl m}\ ,\\
&[Q_{L'}, a_{kl m}] = i(l-m)a_{kl m}\\
&[Q_{\pm \alpha}, a_{kl m}] = -A_{\pm \alpha}(k, l \mp 1, m \mp 1)a_{k, l \mp 1, m\mp 1}\ , \\
&[Q_{\pm \beta}, a_{kl m}] = -A_{\pm \beta}(k, l \pm 1, m \mp 1)a_{k, l \pm 1, m\mp 1}\ ,\\
&[Q_{\pm \gamma}, a_{kl m}]  = -A_{\pm \gamma}(k-1, l \mp 1, m)a_{k-1, l \mp 1, m} - D_{\pm \gamma}(k+1, l \mp 1, m)a_{k+1, l \mp 1, m}\ ,\\
&[Q_{\pm \delta}, a_{kl m}] = -A_{\pm \delta}(k-1, l, m \mp 1)a_{k-1, l , m\mp 1} - D_{\pm \delta}(k+1, l, m \mp 1)a_{k+1, l, m\mp 1}\ .
\end{aligned}
\label{Qacomms}
\end{equation}

The vaccuum is de Sitter invariant, with all $Q| 0\rangle=0$. According to (\ref{qac}), the charges act on
one-particle states (\ref{stdef}) in the following way:
\begin{equation}
\begin{aligned}
&Q_L\big|k\;l\,m\big\rangle  = -i(l+m)\big|k\;l\,m\big\rangle\ ,\\
&Q_{L'}\big|k\;l\,m\big\rangle = -i(l-m)\big|k\;l\,m\big\rangle\ ,\\
&Q_{\pm \alpha}\big|k\;l\,m\big\rangle= A_{\pm \alpha}(k, l, m)\big|k\;l{\pm}1\,m{\pm}1\big\rangle \ ,\\
&Q_{\pm \beta}\big|k\;l\,m\big\rangle = A_{\pm \beta}(k, l, m) \big|k\;l{\mp}1\,m{\pm}1\big\rangle\ ,\\
&Q_{\pm \gamma}\big|k\;l\,m\big\rangle= A_{\pm \gamma}(k, l, m) \big|k{+}1\;l{\pm}1\,m\big\rangle + D_{\pm \gamma}(k, l, m)\big|k{-}1\;l{\pm}1\,m\big\rangle\ ,\\
&Q_{\pm \delta}\big|k\;l\,m\big\rangle = A_{\pm \delta}(k, l, m) \big|k{+}1\;l\,m{\pm}1\big\rangle + D_{\pm \delta}(k, l, m)\big|k{-}1\;l\,m{\pm}1\big\rangle\ .
\end{aligned}\label{stac}
\end{equation}

Since the computations leading to (\ref{stac}), are quite involved, it is important to perform some consistency checks. The first one is to show that one-particle states  (\ref{stdef}) form a {\it unitary\/} representation of de Sitter symmetry group. Indeed, by using the conjugation properties (\ref{cconj}) and the scalar products of (\ref{orthr}), we find
\begin{align}
&Q_L^\dagger = -Q_L\ ,\qquad ~~~ Q_{L'}^\dagger =-Q_{L'}\nonumber\ ,\\
&Q_{\pm\alpha}^\dagger = -Q_{\mp\alpha}\ , \qquad Q_{\pm\beta}^\dagger = -Q_{\mp\beta}\ ,\\
&Q_{\pm\gamma}^\dagger = -Q_{\mp\gamma}\ ,  \qquad Q_{\pm\delta}^\dagger = -Q_{\mp\delta}\  .
\end{align}
Therefore the charges associated to the generators $K_{AB}$ are antihermitean. Acting on one-particle states, they generate unitary transformations. Furthermore, we verified that (\ref{stac}) agree with the commutation relations (\ref{xx1})-(\ref{xx5}). Finally, after identifying  the $SU(2)\times SU(2)'$ quantum numbers as in (\ref{jkrel}),
\be
\j=\jj=\frac{k}{2}\ ,\quad \nu=\frac{l+m}{2}\ ,\quad  \nu'=\frac{l-m}{2}\ ,\qquad\big|k\;l\,m\big\rangle=
\big|(\j\,\nu)(\,\jj\nu')\big\rangle=f^{j,j'}_{\nu,\nu'}\ ,\ee
we find that (\ref{stac}) agree with  Dixmier's results \cite{Dixmier:1961fbm} for the basis of the spin $0$ principal series representation.\footnote{This is after correcting some minor typographical errors in \cite{Dixmier:1961fbm}.}
\begin{figure}[h!]
    \centering
    \begin{tikzpicture}[scale=1.2]
        \draw[step=0.5, thin, color=gray!20] (0,0) grid (5,5);
    
        \draw[->] (0,0) -- (5,0) node[right] {$j$};
        \draw[->] (0,0) -- (0,5) node[above] {$j'$};
    
        \draw[dash pattern=on 3pt off 1pt, line width=0.4pt] (0,0) -- (5,5);
    
        \foreach \k in {0,...,10}{
            \fill (\k*0.5, \k*0.5) circle (2pt);
        }
    
        \node[left] at (0,0.5) {$\frac{1}{2}$};
        \node[below] at (0.5,0) {$\frac{1}{2}$};
        \node[left] at (0,1) {$1$};
        \node[below] at (1,0) {$1$};
    
    \end{tikzpicture}
\caption{Spin 0 principal series representation.}
\end{figure}
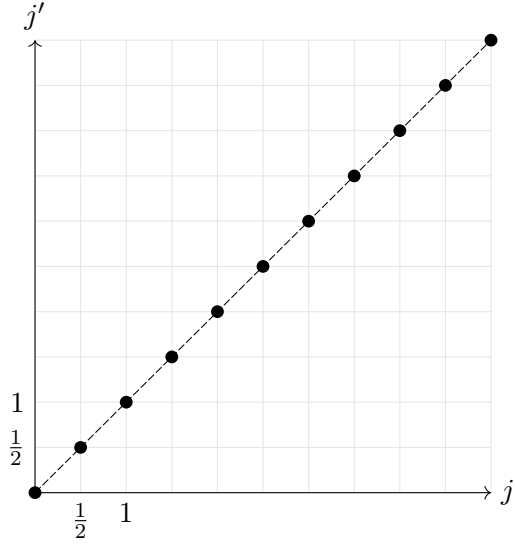

One important comment here is in order. The transformation properties are well-defined not only for $\mu^2\ge 0$ but also for $\mu^2<0$ as long as $\mu^2\ge -\frac{9}{4}$. Although $\mu=0$ is often considered as the boundary between the principal series and the  ``complementary'' series, there is no distinction at the level of transformation properties. The coefficients (\ref{ccoef}) are well-defined and have correct conjugation properties (\ref{cconj}) as long as  $\mu^2\ge -\frac{9}{4}$. The endpoint
$\mu^2= -\frac{9}{4}$ of the complementary series describes a scalar field with $m^2=0$, i.e.\ with zero Lagrangian mass, see (\ref{mkk}). At this point, the coefficients $A_{\pm\gamma}(0,0,0)=A_{\pm\delta}(0,0,0)=0$, and the zero mode state with $k=l=m=0$  becomes a singlet and decouples from the $k\ge 1$ states of the complementary series.

\subsection{Spin $\bf \frac{1}{2}$ principal series} 
The isospin content of spin $ \frac{1}{2}$ principal series is shown in Figure 2. In this case, de Sitter symmetry generators transport the basis states not only along the two lines of $\jj=\j\pm 1/2$  (in NE or SW directions) but also from one line to another (in NW or SE directions). For that reason, it is convenient to label the (anticommuting) creation and annihilation operators as $b_{(\j\,\nu)(\,\jj\nu')}^\dagger$  and $b_{(\j\,\nu)(\,\jj\nu')}$, respectively, and the corresponding one-particle states as
\be\big|(\j\,\nu)(\,\jj\nu')\big\rangle\equiv b_{(\j\,\nu)(\,\jj\nu')}^\dagger\big| 0\big\rangle \ee

One way to determine how the symmetry generators act on one-particle states is to proceed in the same way is in the spin 0 case and examine the 
transformation properties of free spinor fields. It would involve Dirac de Sitter spinors, which are given by rather complicated expressions, therefore we will take a shortcut and start directly from Dixmier's results.

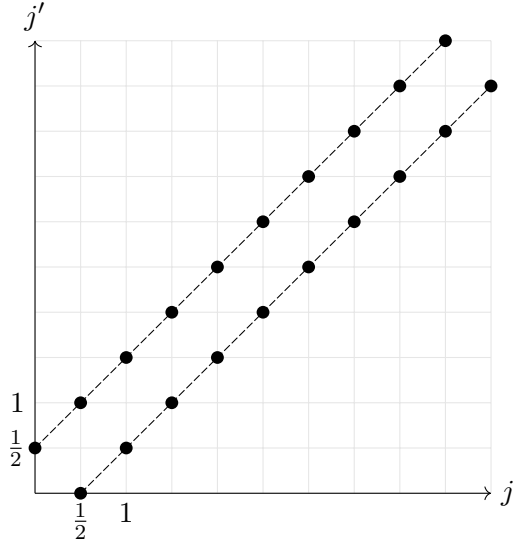
\begin{figure}[t]
\centering
\begin{tikzpicture}[scale=1.2]
\draw[step=0.5, thin, color=gray!20] (0,0) grid (5,5);      
\draw[->] (0,0) -- (5,0) node[right] {$j$};
\draw[->] (0,0) -- (0,5) node[above] {$j'$};
\draw[dash pattern=on 3pt off 1pt, line width=0.4pt] (0,0.5) -- (4.5,5);
\node[left] at (0,0.5) {$\frac{1}{2}$};
\node[below] at (0.5,0) {$\frac{1}{2}$};
\node[left] at (0,1) {$1$};
\node[below] at (1,0) {$1$};
\foreach \k in {0,...,9}{\fill (\k*0.5, \k*0.5 + 0.5) circle (2pt);}
\draw[dash pattern=on 3pt off 1pt, line width=0.4pt] (0.5,0) -- (5, 4.5);
\foreach \k in {0,...,9}{\fill (\k*0.5+0.5, \k*0.5) circle (2pt);} 
\end{tikzpicture}
\caption{Spin $\frac{1}{2}$ principal series representation.}
\end{figure}
We are considering spin $ \frac{1}{2}$ fermions with Lagrangian mass $m=\mu$ in the principal series representation of the de Sitter group. The symmetry generators act on one-particle states in the following way:
\begin{equation}
\begin{aligned}
&Q_L\big|(\j\,\nu)(\,\jj\nu')\big\rangle= -2i\nu\big|(\j\,\nu)(\,\jj\nu')\big\rangle\ ,\\[1mm]
&Q_{L'}\big|(\j\,\nu)(\,\jj\nu')\big\rangle= -2i\nu'\big|(\j\,\nu)(\,\jj\nu')\big\rangle\ ,\\[1mm]
&Q_{\pm \alpha}\big|(\j\,\nu)(\,\jj\nu')\big\rangle= A_{\pm \alpha}(j,j',\nu,\nu')\big|(\j\,\nu{\pm}1)(\,\jj\nu')\big\rangle\ ,\\[1mm]
&Q_{\pm \beta}\big|(\j\,\nu)(\,\jj\nu')\big\rangle= A_{\pm \beta}(j,j',\nu,\nu')\big|(\j\,\nu)(\,\jj\nu'{\mp}1)\big\rangle\ ,\\[1mm]
&Q_{\pm \gamma}\big|(\j\,\nu)(\,\jj\nu')\big\rangle= A_{\pm \gamma}(j,j',\nu,\nu')\textstyle
  \Big|(\j{+}\frac{1}{2}\,\nu{\pm}\frac{1}{2})(\,\jj{+}\frac{1}{2}\,\nu'{\pm}\frac{1}{2})\Big\rangle\\
&\qquad\qquad\qquad~~~~~+B_{\pm \gamma}(j,j',\nu,\nu')\textstyle
  \Big|(\j{-}\frac{1}{2}\,\nu{\pm}\frac{1}{2})(\,\jj{+}\frac{1}{2}\,\nu'{\pm}\frac{1}{2})\Big\rangle\\
&\qquad\qquad\qquad~~~~~+C_{\pm \gamma}(j,j',\nu,\nu')\textstyle
  \Big|(\j{+}\frac{1}{2}\,\nu{\pm}\frac{1}{2})(\,\jj{-}\frac{1}{2}\,\nu'{\pm}\frac{1}{2})\Big\rangle\\
&\qquad\qquad\qquad~~~~~+D_{\pm \gamma}(j,j',\nu,\nu')\textstyle
  \Big|(\j{-}\frac{1}{2}\,\nu{\pm}\frac{1}{2})(\,\jj{-}\frac{1}{2}\,\nu'{\pm}\frac{1}{2})\Big\rangle\ ,\\[1mm]
&Q_{\pm \delta} \big|(\j\,\nu)(\,\jj\nu')\big\rangle= 
A_{\pm \delta}(j,j',\nu,\nu')\textstyle
  \Big|(\j{+}\frac{1}{2}\,\nu{\pm}\frac{1}{2})(\,\jj{+}\frac{1}{2}\,\nu'{\mp}\frac{1}{2})\Big\rangle\\
&\qquad\qquad\qquad~~~~~+B_{\pm \delta}(j,j',\nu,\nu')\textstyle
  \Big|(\j{-}\frac{1}{2}\,\nu{\pm}\frac{1}{2})(\,\jj{+}\frac{1}{2}\,\nu'{\mp}\frac{1}{2})\Big\rangle\\
&\qquad\qquad\qquad~~~~~+C_{\pm \delta}(j,j',\nu,\nu')\textstyle
  \Big|(\j{+}\frac{1}{2}\,\nu{\pm}\frac{1}{2})(\,\jj{-}\frac{1}{2}\,\nu'{\mp}\frac{1}{2})\Big\rangle\\
&\qquad\qquad\qquad~~~~~+D_{\pm \delta}(j,j',\nu,\nu')\textstyle
  \Big|(\j{-}\frac{1}{2}\,\nu{\pm}\frac{1}{2})(\,\jj{-}\frac{1}{2}\,\nu'{\mp}\frac{1}{2})\Big\rangle\ ,
\end{aligned}\label{stac1}
\end{equation}
with the coefficients
\begin{equation}
\begin{aligned}
A_{\pm \alpha}(j, j', \nu,\nu')  &= \pm 2\sqrt{(j \pm \nu + 1)(j\mp \nu)}~\qquad\qquad\qquad~\quad~~~~~~~\qquad\qquad~~~~~~~~~~~\\[2mm]
A_{\pm \beta}(j, j', \nu,\nu') & = \pm 2 \sqrt{(j' \mp \nu' + 1)(j' \pm \nu')}
\end{aligned}\label{fc1}
\end{equation}
\begin{equation}
\begin{aligned}
A_{\pm \gamma}(j, j', \nu, \nu') & = -i \sqrt{\frac{(j\pm\nu+1)(j'\pm \nu' +1)(j+j'+\frac{1}{2})(j+j'+\frac{3}{2})((j+j' + \frac{3}{2})^2+\mu^2)}{(2j+1)(2j+2)(2j'+1)(2j'+2)}}\\[1mm]
B_{\pm \gamma}(j, j', \nu, \nu') & = \mp i \sqrt{\frac{(j\mp \nu)(j'\pm \nu'+1)(j-j'-\frac{3}{2})(j-j'+\frac{1}{2})((j-j'-\frac{1}{2})^2+\mu^2)}{(2j)(2j+1)(2j'+1)(2j'+2)}}\\[1mm]
C_{\pm \gamma}(j, j', \nu, \nu') & = \pm i \sqrt{\frac{(j'\mp \nu')(j\pm \nu+1)(j-j'-\frac{1}{2})(j-j'+\frac{3}{2})((j-j'+\frac{1}{2})^2+\mu^2)}{(2j+1)(2j+2)(2j')(2j'+1)}}\\[1mm]
D_{\pm \gamma}(j, j', \nu, \nu') & = -i \sqrt{\frac{(j\mp \nu)(j'\mp\nu')(j+j'-\frac{1}{2})(j+j'+\frac{1}{2})((j+j'+\frac{1}{2})^2+\mu^2)}{(2j)(2j+1)(2j')(2j'+1)}}\\[1mm]
\end{aligned}\label{fc2}
\end{equation}
\begin{equation}
\begin{aligned}
A_{\pm \delta}(j, j', \nu, \nu') & = \mp i \sqrt{\frac{(j\pm\nu +1)(j'\mp\nu' +1)(j+j'+\frac{1}{2})(j+j'+\frac{3}{2})((j+j' + \frac{3}{2})^2+\mu^2)}{(2j+1)(2j+2)(2j'+1)(2j'+2)}}\\[1mm]
B_{\pm \delta}(j, j', \nu, \nu') & = i \sqrt{\frac{(j\mp \nu)(j'\mp \nu'+1)(j-j'-\frac{3}{2})(j-j'+\frac{1}{2})((j-j'-\frac{1}{2})^2+\mu^2)}{(2j)(2j+1)(2j'+1)(2j'+2)}}\\[1mm]
C_{\pm \delta}(j, j', \nu, \nu') & = i \sqrt{\frac{(j\pm \nu+1)(j'\pm \nu')(j-j'-\frac{1}{2})(j-j'+\frac{3}{2})((j-j'+\frac{1}{2})^2+\mu^2)}{(2j+1)(2j+2)(2j')(2j'+1)}}\\[1mm]
D_{\pm \delta}(j, j', \nu, \nu') & = \pm i \sqrt{\frac{(j\mp \nu)(j'\pm\nu')(j+j'-\frac{1}{2})(j+j'+\frac{1}{2})((j+j'+\frac{1}{2})^2+\mu^2)}{(2j)(2j+1)(2j')(2j'+1)}}
\end{aligned}\label{fc3}
\end{equation}

Upon complex conjugation,
\begin{equation}
\begin{aligned}
A^{\dagger}_{\pm \alpha}(j, j', \nu, \nu')& = A_{\pm \alpha}(j, j', \nu, \nu') = -A_{\mp \alpha}(j, j', \nu \mp 1, \nu') \qquad\qquad\qquad~~\\[1mm]
A^{\dagger}_{\pm \beta}(j, j', \nu, \nu')& = A_{\pm \beta}(j, j', \nu, \nu') = -A_{\mp \beta}(j, j', \nu, \nu' \pm 1) 
\end{aligned}\label{conj1}
\end{equation}
\begin{equation}
\begin{aligned}
A^{\dagger}_{\pm \gamma}(j, j', \nu, \nu')& = -A_{\pm \gamma}(j, j', \nu, \nu') = -D_{\mp \gamma}(j + \tfrac{1}{2}, j' + \tfrac{1}{2}, \nu \mp \tfrac{1}{2}, \nu' \mp \tfrac{1}{2}) \\
 B^{\dagger}_{\pm \gamma}(j, j', \nu, \nu') & = -B_{\pm \gamma}(j, j', \nu, \nu')=-C_{\mp \gamma}(j-\tfrac{1}{2}, j+\tfrac{1}{2}, \nu \mp \tfrac{1}{2}, \nu' \mp \tfrac{1}{2})\\
C^{\dagger}_{\pm \gamma}(j, j', \nu, \nu')& = -C_{\pm \gamma}(j, j', \nu, \nu') = -B_{\mp \gamma}(j+\tfrac{1}{2}, j'-\tfrac{1}{2}, \nu \mp \tfrac{1}{2}, \nu' \mp \tfrac{1}{2}) \\ 
D_{\pm \gamma}^{\dagger}(j, j', \nu, \nu')& = -D_{\pm \gamma}(j, j', \nu, \nu') = -A_{\mp \gamma}(j -\tfrac{1}{2}, j' - \tfrac{1}{2}, \nu \mp \tfrac{1}{2}, \nu' \mp \tfrac{1}{2}) 
\end{aligned}\label{conj2}
\end{equation}

\begin{equation}
\begin{aligned}
A^{\dagger}_{\pm \delta}(j, j', \nu, \nu') & = -A_{\pm \delta}(j, j', \nu, \nu') = -D_{\mp \delta}(j + \tfrac{1}{2}, j' + \tfrac{1}{2}, \nu \mp \tfrac{1}{2}, \nu' \pm \tfrac{1}{2})\\
B^{\dagger}_{\pm \delta}(j, j', \nu, \nu') & = -B_{\pm \delta}(j, j', \nu, \nu') =- C_{\mp \delta}(j-\tfrac{1}{2}, j+\tfrac{1}{2}, \nu \mp \tfrac{1}{2}, \nu' \pm \tfrac{1}{2})\\
 C^{\dagger}_{\pm \delta}(j, j', \nu, \nu')& = -C_{\pm \delta}(j, j', \nu, \nu') =-B_{\mp \delta}(j+\tfrac{1}{2}, j'-\tfrac{1}{2}, \nu \mp \tfrac{1}{2}, \nu' \pm \tfrac{1}{2}) \\ 
D_{\pm \delta}^{\dagger}(j, j', \nu, \nu')& = -D_{\pm \delta}(j, j', \nu, \nu') = -A_{\mp \delta}(j -\tfrac{1}{2}, j' - \tfrac{1}{2}, \nu \mp \tfrac{1}{2}, \nu' \pm \tfrac{1}{2}) \end{aligned}\label{conj3}
\end{equation}
By using the above conjugation properties, we verified that the representation is unitary. We also checked that the coefficients written in (\ref{fc1})-(\ref{fc3}) agree with the symmetry algebra.
The commutation relations of the symmetry generators with the creation operators are written below:
\begin{equation}
\begin{aligned}
&[Q_L,b^\dagger_{(\j\,\nu)(\,\jj\nu')}]= -2i\nu \,b^\dagger_{(\j\,\nu)(\,\jj\nu')}\ ,\\[1mm]
&[Q_{L'},b^\dagger_{(\j\,\nu)(\,\jj\nu')}]= -2i\nu'\,b^\dagger_{(\j\,\nu)(\,\jj\nu')}\ ,\\[1mm]
&[Q_{\pm \alpha},b^\dagger_{(\j\,\nu)(\,\jj\nu')}]= A_{\pm \alpha}(j,j',\nu,\nu')\, b^\dagger_{(\j\,\nu{\pm}1)(\,\jj\nu')}\ ,\\[1mm]
&[Q_{\pm \beta},b^\dagger_{(\j\,\nu)(\,\jj\nu')}]= A_{\pm \beta}(j,j',\nu,\nu')\, b^\dagger_{(\j\,\nu)(\,\jj\nu'{\mp}1)}\ ,\\[1mm]
&[Q_{\pm \gamma},b^\dagger_{(\j\,\nu)(\,\jj\nu')}]= A_{\pm \gamma}(j,j',\nu,\nu')\textstyle
 \, b^\dagger_{(\j{+}\frac{1}{2}\,\nu{\pm}\frac{1}{2})(\,\jj{+}\frac{1}{2}\,\nu'{\pm}\frac{1}{2})}\\
&\qquad\qquad\qquad~~~~~+B_{\pm \gamma}(j,j',\nu,\nu')\textstyle
 \, b^{\dagger}_{(\j{-}\frac{1}{2}\,\nu{\pm}\frac{1}{2})(\,\jj{+}\frac{1}{2}\,\nu'{\pm}\frac{1}{2})}\\
&\qquad\qquad\qquad~~~~~+C_{\pm \gamma}(j,j',\nu,\nu')\textstyle
 \, b^{\dagger}_{(\j{+}\frac{1}{2}\,\nu{\pm}\frac{1}{2})(\,\jj{-}\frac{1}{2}\,\nu'{\pm}\frac{1}{2})}\\
&\qquad\qquad\qquad~~~~~+D_{\pm \gamma}(j,j',\nu,\nu')\textstyle
 \, b^{\dagger}_{(\j{-}\frac{1}{2}\,\nu{\pm}\frac{1}{2})(\,\jj{-}\frac{1}{2}\,\nu'{\pm}\frac{1}{2})}\ ,\\[1mm]
&[Q_{\pm \delta} ,b^\dagger_{(\j\,\nu)(\,\jj\nu')}]= 
A_{\pm \delta}(j,j',\nu,\nu')\textstyle
 \, b^{\dagger}_{(\j{+}\frac{1}{2}\,\nu{\pm}\frac{1}{2})(\,\jj{+}\frac{1}{2}\,\nu'{\mp}\frac{1}{2})}\\
&\qquad\qquad\qquad~~~~~+B_{\pm \delta}(j,j',\nu,\nu')\textstyle
  \, b^{\dagger}_{(\j{-}\frac{1}{2}\,\nu{\pm}\frac{1}{2})(\,\jj{+}\frac{1}{2}\,\nu'{\mp}\frac{1}{2})}\\
&\qquad\qquad\qquad~~~~~+C_{\pm \delta}(j,j',\nu,\nu')\textstyle
  \, b^{\dagger}_{(\j{+}\frac{1}{2}\,\nu{\pm}\frac{1}{2})(\,\jj{-}\frac{1}{2}\,\nu'{\mp}\frac{1}{2})}\\
&\qquad\qquad\qquad~~~~~+D_{\pm \delta}(j,j',\nu,\nu')\textstyle
 \, b^{\dagger}_{(\j{-}\frac{1}{2}\,\nu{\pm}\frac{1}{2})(\,\jj{-}\frac{1}{2}\,\nu'{\mp}\frac{1}{2})}\ . \end{aligned}\label{bcom1}
\end{equation} 
Similarly, the commutation relations of the symmetry generators with the annihilation operators are:
\begin{equation}
\begin{aligned}
&[Q_L,b_{(\j\,\nu)(\,\jj\nu')}]= 2i\nu \,b_{(\j\,\nu)(\,\jj\nu')}\ ,\\[1mm]
&[Q_{L'},b_{(\j\,\nu)(\,\jj\nu')}]= 2i\nu'\,b_{(\j\,\nu)(\,\jj\nu')}\ ,\\[1mm]
&[Q_{\pm \alpha}, b_{(\j\,\nu)(\,\jj\nu')}] 
=  -A_{\pm \alpha}(j,j',\nu\mp 1,\nu')\, b_{(\j\,\nu{\mp}1)(\,\jj\nu')}\ ,\\[1mm]
&[Q_{\pm \beta}, b_{(\j\,\nu)(\,\jj\nu')}] 
= -A_{\pm \beta}(j,j',\nu,\nu'\pm 1)\, b_{(\j\,\nu)(\,\jj\,\nu'{\pm}1)}\ ,\\[1mm]
&[Q_{\pm \gamma},b_{(\j\,\nu)(\,\jj\nu')}]=\,\textstyle -A_{\pm \gamma}(j{-}\frac{1}{2},j'{-}\frac{1}{2},\nu{\mp}\frac{1}{2},\nu'{\mp}\frac{1}{2})\textstyle
 \, b_{(\j{-}\frac{1}{2}\,\nu{\mp}\frac{1}{2})(\,\jj{-}\frac{1}{2}\,\nu'{\mp}\frac{1}{2})}\\
&\qquad\qquad\qquad~~~~~~\textstyle - B_{\pm \gamma}(j{+}\frac{1}{2},j'{-}\frac{1}{2},\nu{\mp}\frac{1}{2},\nu'{\mp}\frac{1}{2})\textstyle
 \, b_{(\j{+}\frac{1}{2}\,\nu{\mp}\frac{1}{2})(\,\jj{-}\frac{1}{2}\,\nu'{\mp}\frac{1}{2})}\\
&\qquad\qquad\qquad~~~~~~\textstyle -C_{\pm \gamma}(j{-}\frac{1}{2},j'{+}\frac{1}{2},\nu{\mp}\frac{1}{2},\nu'{\mp}\frac{1}{2})\textstyle
 \, b_{(\j{-}\frac{1}{2}\,\nu{\mp}\frac{1}{2})(\,\jj{+}\frac{1}{2}\,\nu'{\mp}\frac{1}{2})}\\
&\qquad\qquad\qquad~~~~~~\textstyle -D_{\pm \gamma}(j{+}\frac{1}{2},j'{+}\frac{1}{2},\nu{\mp}\frac{1}{2},\nu'{\mp}\frac{1}{2})\textstyle
 \, b_{(\j{+}\frac{1}{2}\,\nu{\mp}\frac{1}{2})(\,\jj{+}\frac{1}{2}\,\nu'{\mp}\frac{1}{2})}\ ,\\[1mm]
&[Q_{\pm \delta},b_{(\j\,\nu)(\,\jj\nu')}]=\,\textstyle -A_{\pm \delta}(j{-}\frac{1}{2},j'{-}\frac{1}{2},\nu{\mp}\frac{1}{2},\nu'{\pm}\frac{1}{2})\textstyle
 \, b_{(\j{-}\frac{1}{2}\,\nu{\mp}\frac{1}{2})(\,\jj{-}\frac{1}{2}\,\nu'{\pm}\frac{1}{2})}\\
&\qquad\qquad\qquad~~~~~~\textstyle - B_{\pm \delta}(j{+}\frac{1}{2},j'{-}\frac{1}{2},\nu{\mp}\frac{1}{2},\nu'{\pm}\frac{1}{2})\textstyle
 \, b_{(\j{+}\frac{1}{2}\,\nu{\mp}\frac{1}{2})(\,\jj{-}\frac{1}{2}\,\nu'{\pm}\frac{1}{2})}\\
&\qquad\qquad\qquad~~~~~~\textstyle -C_{\pm \delta}(j{-}\frac{1}{2},j'{+}\frac{1}{2},\nu{\mp}\frac{1}{2},\nu'{\pm}\frac{1}{2})\textstyle
 \, b_{(\j{-}\frac{1}{2}\,\nu{\mp}\frac{1}{2})(\,\jj{+}\frac{1}{2}\,\nu'{\pm}\frac{1}{2})}\\
&\qquad\qquad\qquad~~~~~~\textstyle -D_{\pm \delta}(j{+}\frac{1}{2},j'{+}\frac{1}{2},\nu{\mp}\frac{1}{2},\nu'{\pm}\frac{1}{2})\textstyle
 \, b_{(\j{+}\frac{1}{2}\,\nu{\mp}\frac{1}{2})(\,\jj{+}\frac{1}{2}\,\nu'{\pm}\frac{1}{2})}\ .
\end{aligned}\label{bcom2}
\end{equation}

In the zero mass limit $m=\mu=0$ , the coefficients $B$ and $C$ vanish for the entire representation. As a result, the generators $Q_{\pm \gamma}$ and $Q_{\pm \delta}$ no longer mix the two branches with $\jj - \j = \pm \tfrac12$, see Figure 2. The principal series representation therefore splits into two ``discrete'' UIRs of chiral fermions: $\Pi^+$ with helicity $\jj-\j=\frac{1}{2}$ and $\Pi^-$ with helicity $\jj-\j=-\frac{1}{2}$. Their transformation properties and the commutators of creation/annihilation operators with the symmetry generators can be obtained from the above expressions by setting $\mu=0$ and restricting to $\jj=\j\pm\frac{1}{2}$
for $\Pi^+$ and $\Pi^-$, respectively.

\subsection{Spin 1 gauge bosons and spin 2 gravitons}

\begin{figure}[h]
\centering
\begin{tikzpicture}[scale=1.2]
\draw[step=0.5, thin, color=gray!20] (0,0) grid (5,5);
\draw[->] (0,0) -- (5,0) node[right] {$j$};
\draw[->] (0,0) -- (0,5) node[above] {$j'$};
\draw[dash pattern=on 3pt off 1pt, line width=0.4pt, red] (0,1) -- (4,5);
\draw[dash pattern =on 3pt off 1pt, line width=0.4pt, blue] (1,0) -- (5, 4);
\node[left] at (0,1) {$1$};
\node[below] at (1,0) {$1$};
\foreach \k in {0,...,8}{\fill[red] (\k*0.5, \k*0.5 + 1) circle (2pt);}
\foreach \k in {0,...,8}{\fill[blue] (\k*0.5+1, \k*0.5) circle (2pt);}
\node[red] at (2,3.5) {${\Pi^+_1}$}; 
\node[blue] at (3.5,2) {${\Pi^-_1}$};
\node[left] at (0,0.5) {$\frac{1}{2}$};
\node[below] at (0.5,0) {$\frac{1}{2}$};
\node[left] at (0,1) {$1$};
\node[below] at (1,0) {$1$};
\end{tikzpicture}
\caption{Spin 1 representations for polarized gauge bosons.}
\end{figure}
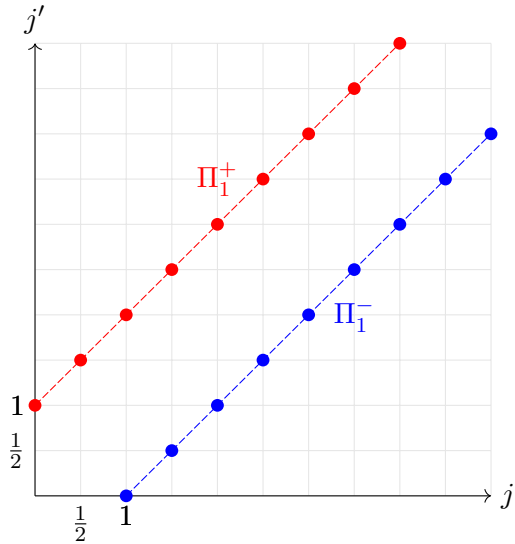
\begin{figure}[h!]
\centering
\begin{tikzpicture}[scale=1.2]
\draw[step=0.5, thin, color=gray!20] (0,0) grid (5,5);
\draw[->] (0,0) -- (5,0) node[right] {$j$};
\draw[->] (0,0) -- (0,5) node[above] {$j'$};
\draw[dash pattern=on 3pt off 1pt, line width=0.4pt, red] (0,2) -- (3,5);
\draw[dash pattern =on 3pt off 1pt, line width=0.4pt, blue] (2,0) -- (5, 3);
\node[left] at (0,2) {$2$};
\node[below] at (2,0) {$2$};
\foreach \k in {0,...,6}{\fill[red] (\k*0.5, \k*0.5 + 2) circle (2pt);}
\foreach \k in {0,...,6}{\fill[blue] (\k*0.5+2, \k*0.5) circle (2pt);}
\node[red] at (1.5,4) {${\Pi^+_2}$};
\node[blue] at (4,1.5) {${\Pi^-_2}$};
\node[left] at (0,0.5) {$\frac{1}{2}$};
\node[below] at (0.5,0) {$\frac{1}{2}$};
\node[left] at (0,1) {$1$};
\node[below] at (1,0) {$1$};
\node[left] at (0,1.5) {$\frac{3}{2}$};
\node[below] at (1.5,0) {$\frac{3}{2}$};
\end{tikzpicture}
\caption{Spin 2 representations for polarized gravitons}
\end{figure}
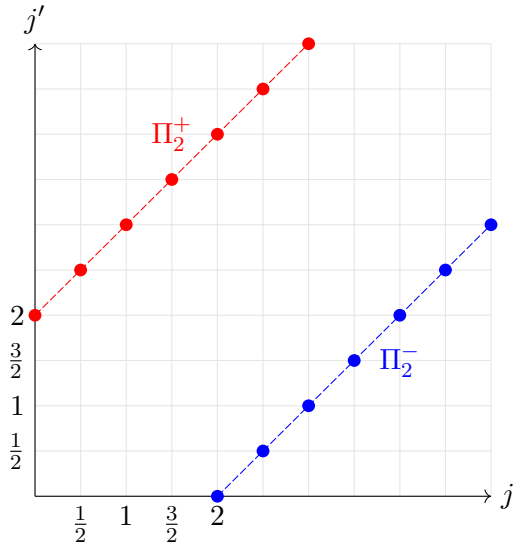
The isospin contents of two UIRs representing spin 1 gauge bosons, $\Pi^+_1$($\jj=\j+1)$ for right-handed polarizations and $\Pi^-_1$($\jj=\j-1$) for left-handed polarizations, are shown in Figure 3. The isospin content of two UIRs representing spin 2 gravitons,  $\Pi^+_2$($\jj=\j+2)$ for right-handed polarizations and $\Pi^-_2$($\jj=\j-2$) for left-handed polarizations, are shown in Figure 4. They belong to the ``discrete'' family of representations.
It is convenient to label the states as in section 4.2, by 
\be\big|(\j\,\nu)(\,\jj\nu')\rangle\equiv b_{(\j\,\nu)(\,\jj\nu')}^\dagger\big| 0\big\rangle 
~~\makebox{with}~~ \begin{cases}~\jj=\j+1~~\makebox{for}~~\Pi^+_1\\ ~\jj=\j-1~~\makebox{for}~~\Pi^-_1\\
~\jj=\j+2~~\makebox{for}~~\Pi^+_2\\ ~\jj=\j-2~~\makebox{for}~~\Pi^-_2
\end{cases}\ee
An interesting feature of these representations is the absence of ``soft'' particles with the ``energies'' $k=\j+\jj=0$.\footnote {We use the term ``energy'' to characterize $k$, anticipating its physical interpretation in the flat limit as described in the next section. Its group-theoretic meaning, however, is different. For example, in the case of scalar principal series, $k$ corresponds to the $SO(4)$ spin: the rank of the symmetric traceless tensor representation of the $SO(4)$ subgroup of   $SO(1,4)$  \cite{rep4}.} For gauge bosons, the spectrum starts at $k=1$, while for gravitons at $k=2$.

The gauge boson and graviton representations share the same transformation properties.  The symmetry generators act on one-particle states as in (\ref{stac1}),
with the coefficients $B=C=0$ and
\begin{equation}
\begin{aligned}
&A_{\pm \alpha}(j, j', \nu, \nu') = \pm 2\sqrt{(j \pm \nu + 1)(j\mp \nu)}\\
&A_{\pm \beta}(j, j', \nu, \nu') = \pm 2 \sqrt{(j' \mp \nu' + 1)(j' \pm \nu')}\\[1mm]
&A_{\pm \gamma}(j, j', \nu, \nu')  = -i\sqrt{(j\pm\nu+1)(j'\pm\nu'+1)}\\
&D_{\pm \gamma}(j, j', \nu, \nu')  = -i\sqrt{(j\mp \nu)(j'\mp\nu')}\\[1mm]
&A_{\pm \delta}(j, j', \nu, \nu')  = \mp i \sqrt{(j\pm\nu +1)(j' \mp \nu' +1)}\\
&D_{\pm \delta}(j, j', \nu, \nu')  = \pm i \sqrt{(j \mp \nu)(j' \pm \nu')}
\end{aligned}\label{acco}
\end{equation}
Their conjugation properties are listed in (\ref{conj1})-(\ref{conj3}). Furthermore, the commutators of the symmetry generators with the creation and annihilation operators are exactly the same as in (\ref{bcom1}) and (\ref{bcom2}), now specified to the coefficients written in (\ref{acco}).

\section{Ward identities}
\subsection{General structure}
One of the fundamental assumptions of quantum field theory is the existence of a unitary time evolution operator $U(\tau,\tau')$ which acts on Hilbert space $\cal H$, evolving quantum states from time $\tau'$ to $\tau$.  In the operator formulation, the transition amplitude from  a freely propagating  initial $|\bm i\rangle$ state, as observed in the asymptotic past, to a free  final state $|\bm f\rangle$, as observed in the asymptotic future (with both initial and final states far from the interaction region) is given by
\be{\cal M}_{fi}=\langle\bm{f}|U(\tau=\infty,\tau'=-\infty)|\bm{i}\rangle\equiv\langle\bm{f}|U|\bm{i}     \rangle\ee
In the case of de Sitter spacetime, it is natural  to express the initial and final states in the basis of ${\cal H}= \oplus_{(j, j')\in \Gamma}{\cal H}_{j,j'}$. In \cite{Taylor:2024vdc}, the operator $U$ was given  in Dyson's form
\be U=T\, \makebox{exp}\Big(-i\int d^dx
\sqrt{-g}\,H_I[\phi(x)]
\Big),\label{dys}\ee
where $H_I$ is the interaction Hamiltonian and the time ordering is with respect to the time coordinate of the embedding Minkowski spacetime.
Here, however, we will not use this particular form. The most important, universal  property of the time evolution operator is its invariance under the symmetries of  interactions. For all conserved charges,
\be [Q, U]=0.\label{vin}\ee

We are considering $N$-particle amplitudes
\begin{align}
\big\langle(\scriptstyle k_{r+1},&\scriptstyle l_{r+1},m_{r+1})_{P_{r+1}}\dots (k_N,l_N,m_N)_{P_N}\big|(k_1,l_1,m_1)_{P_1}\dots (k_r,l_r,m_r)_{P_r}\big\rangle\nonumber\\[2mm] &=~\langle 0|a^{P_{r+1}}_{k_{r+1}l_{r+1}m_{r+1}}\cdots a^{P_{N}}_{k_{N}l_{N}m_{N}}\, U\, a^{P_1\dagger}_{k_1l_1 m_1}\cdots a^{P_r\dagger}_{k_rl_rm_r}|0\rangle\,,
\end{align}
where $P_i$ denotes the particle species. Ward identities follow from de Sitter invariance of the vacuum, $Q|0\rangle=0$, and from the symmetry of the time evolution operator, (\ref{vin}). For any de Sitter symmetry generator,
\begin{align}
&0~=~\langle 0|\,[\,Q,a^{P_{r+1}}_{k_{r+1}l_{r+1}m_{r+1}}\cdots a^{P_{N}}_{k_{N}l_{N}m_{N}}\, U\, a^{P_1\dagger}_{k_1l_1 m_1}\cdots a^{P_r\dagger}_{k_rl_rm_r}]\,|0\rangle\\
&=\sum_{i=1}^r\sum_{k'_i,l'_i,m'_i}\!Q_{ k_i,l_i,m_i}^{k'_i,l'_i,m'_i}\,\big\langle\scriptstyle( k_{r+1},\scriptstyle l_{r+1},m_{r+1})_{P_{r+1}}\dots (k_N,l_N,m_N)_{P_N}\big|(k_1,l_1,m_1)_{P_1}\dots  (k'_i,l'_i,m'_i)_{P_i}\dots (k_r,l_r,m_r)_{P_r}\big\rangle\nonumber\\
&~+\!\!\sum_{i=r+1}^N\sum_{k'_i,l'_i,m'_i}\!\overline{Q}_{ k_i,l_i,m_i}^{k'_i,l'_i,m'_i}\,\big\langle(\scriptstyle k_{r+1},\scriptstyle l_{r+1},m_{r+1})_{P_{r+1}}\dots (k'_i,l'_i,m'_i)_{P_i}\dots (k_N,l_N,m_N)_{P_N}\big|(k_1,l_1,m_1)_{P_1}\dots  (k_r,l_r,m_r)_{P_r}\big\rangle,
\nonumber
\end{align}
with the coefficient determined by the commutators
\begin{align}
[Q,  a^{P\dagger}_{kl m}]&=\sum_{k',l',m'}Q_{ k,l,m}^{k',l',m'}a^{P\dagger }_{k'l'm'}\ ,\\
[Q,   a^{P}_{kl m}]&=\sum_{k',l',m'}\overline{Q}_{ k,l,m}^{k',l',m'}a^{P}_{k'l'm'}\  ,\end{align}
which, for all particles $P$ of the standard model and gravitons, can be read off from the results of section \nolinebreak 4.

The Ward identities associated to the $SU(2)\times SU(2)'$ isospin charges $Q_L, Q_{\pm\alpha}$ and $Q_{L'}, Q_{\pm\beta}$, respectively, are particularly simple. From $Q_L$ and $Q_{L'}$, we obtain the con{\nolinebreak}serva{\nolinebreak}tion laws
\be \sum_{i=1}^rl_i-\!\sum_{i=r+1}^Nl_i=0\ ,\qquad \sum_{i=1}^rm_i-\!\sum_{i=r+1}^Nm_i=0\ ,\label{lmcons}\ee
while  $Q_{\pm\alpha}$ and $Q_{\pm\beta}$ act as the ladder operators of $SU(2)$ and $SU(2)'$, respectively, and generate standard isospin Ward identities. They imply standard selection rules, which become transparent after rewriting (\ref{lmcons}) as
\be \sum_{i=1}^r\nu_i-\!\sum_{i=r+1}^N\nu_i=0\ ,\qquad \sum_{i=1}^r\nu'_i-\!\sum_{i=r+1}^N\nu'_i=0\ ,\label{nucons}\ee 
with $\nu$ and $\nu'$ taking half-integer or integer values, depending on $\j$ and $\jj$, see (\ref{dst}). This is possible only if
\be \sum_{i=1}^N\j_i\in\mathbb{Z}\ , \qquad \sum_{i=1}^N\jj_i\in\mathbb{Z}\ .\label{intc}\ee
The above constraint, together with the conditions
\be \sup\{\j_i\}\le \frac{1}{2}\sum_{i=1}^N\j_i\ ,\qquad  \sup\{\jj_i\}\le \frac{1}{2}\sum_{i=1}^N\jj_i  \ee
are necessary  for the  isospin invariance of amplitudes.

On the other hand, the quantum numbers $\j_i$ and $\jj_i$ are related to $k_i$ and to the helicity $s_i$ of particle $P_i$ through
\be \j_i+ \jj_i=k_i, \quad \jj_i-\j_i= s_i\ ,\ee
where $s_i$ may be integer or half-integer depending on the particle species. It follows that
\be \sum_{i=1}^{N}k_i=\frac{1}{2}[N(s=+1/2)-N(s=-1/2)]+N(s=+1)-N(s=-1)~~\makebox{mod}~~2.\ee

The Ward identities associated with $Q_{\pm \gamma}$ and $Q_{\pm \delta}$ link particles belonging to different isospin multiplets. They can lead to some interesting relations between ``soft'' amplitudes with low $k$ and higher (angular) momentum amplitudes, as well as to ``crossing'' relations, as illustrated in the examples discussed below.
\subsection{Examples}
To illustrate the consequences of Ward identities associated with the isospin-changing $Q_{\pm \gamma}$ and $Q_{\pm \delta}$, we focus on the scalar principal series. In this case, the coefficients $Q_{ k_i,l_i,m_i}^{k'_i,l'_i,m'_i}$ can be read from (\ref{ccoef}).

We are particularly interested in the processes in which the vacuum ``radiates away'' a number of particles -- the processes that violate energy-momentum conservation in Minkowski spacetime and would normally signal a vacuum instability. As pointed out in \cite{Taylor:2025deepIR}, they are possible in a de Sitter background. Let us consider three-scalar ``all out'' amplitudes, beginning with the amplitude for the creation of three isospin singlets with $(\j,\jj)=(0,0)$, {\it i.e}.\ $k_1=k_2=k_3=0$. We can relate it to other amplitudes by using
\begin{align}
\langle 0|\,[\,Q_{-\gamma},a_{1{-}10}a_{000}a_{000}\, U\,]\,|0\rangle&=0=\big\langle\scriptstyle (000)(000)(000)\big|0 \big\rangle {\displaystyle \,+\,2}\big\langle\scriptstyle (1{-}10)(110)(000)\big|0 \big\rangle\ ,\\[1mm]
\langle 0|\,[\,Q_{-\delta},a_{10{-}1}a_{000}a_{000}\, U\,]\,|0\rangle&=0=\big\langle\scriptstyle (000)(000)(000)\big|0 \big\rangle {\displaystyle \,-\,2}\big\langle\scriptstyle (10{-}1)(101)(000)\big|0 \big\rangle\ .\end{align}
On the other hand, the Ward identities due to the $Q_{\pm\alpha}$ and $Q_{\pm\beta}$ isospin generators yield
\be \big\langle\scriptstyle (10{-}1)(101)(000)\big|0 \big\rangle {\displaystyle~=~-}\big\langle\scriptstyle (1{-}10)(110)(000)\big|0 \big\rangle\ ,
\ee
therefore
\be \big\langle\scriptstyle (10{-}1)(101)(000)\big|0 \big\rangle {\displaystyle ~=~-}\big\langle\scriptstyle (1{-}10)(110)(000)\big|0 \big\rangle {\displaystyle~=\frac{1}{2}}\big\langle\scriptstyle (000)(000)(000)\big|0 \big\rangle\ .
\ee
In the same way, we can show that
\be
\big\langle\scriptstyle (200)(1{-}10)(110)\big|0 \big\rangle {\displaystyle~=~}{\textstyle \frac{1}{2\sqrt 3}}
\sqrt{\frac{\mu^2+\frac{9}{4}}{\mu^2+\frac{25}{4}}} 
\big\langle\scriptstyle (000)(000)(000)\big|0 \big\rangle\ .\label{ex1}
\ee
and relate it to other amplitudes involving one $k=2$ and two $k=1$ particles using isospin symmetry. Up to this point, all amplitudes can be expressed in terms of the ``zero mode'' amplitude with three $k=0$ particles. This is no longer possible, however, for the amplitudes involving more particles with $k=2$ or for the amplitudes with $k>2$ because their number  exceeds the number of constraints imposed by Ward identities. For example, starting from
\be \langle 0|\,[\,Q_{-\gamma},a_{1{-}10}a_{200}a_{200}\, U\,]\,|0\rangle=0\ ,\ee
we obtain
\begin{align}2{\textstyle\sqrt{\frac{\mu^2 + \frac{25}{4}}{6}} }\big\langle{\scriptstyle(200)(1{-}10)(110)\big|0 \big\rangle}~+~& {\textstyle\sqrt{\frac{\mu^2+\frac{9}{4}}{2}}}\big\langle{\scriptstyle(200)(200)(000)\big|0 \big\rangle}\nonumber
\\[1mm]
+\, {\textstyle \sqrt{\frac{\mu^2+\frac{25}{4}}{6}} }\big\langle{\scriptstyle(200)(200)(200)\big|0 \big\rangle}&
+2{\textstyle\sqrt{\frac{\mu^2+ \frac{49}{4}}{3}}}\big\langle{\scriptstyle(310)(200)(1{-}10)\big|0 \big\rangle}=0\ .\label{ex2}\end{align}
While the first amplitude on the  l.h.s.\ can be expressed in terms of the zero mode amplitude by using (\ref{ex1}), only a specific combination of the remaining terms is determined by the above Ward identity. It is easy to see that other Ward identities yield either isospin-transformed versions of (\ref{ex2}) or they involve amplitudes with $k>3$. This is not unexpected because the symmetries of the underlying theory are not expected to determine the scattering amplitudes.

There is, however, an interesting conclusion that can be drawn from the above discussion. In \cite{Taylor:2025deepIR}  the zero mode amplitude $\langle{\scriptstyle (000)(000)(000)\big|0 }\rangle$ was found to be zero in the interacting $\phi^3$ theory. This is related to the fact that the tensor product of three principal series does not contain the trivial representation\footnote{We thank the referee for pointing this out.} \cite{rep4} in the following way. A non-singlet state cannot be produced from the vacuum without violating the symmetry. For finite-dimensional representations of compact symmetry groups, Ward identities yield linear combinations of non-singlet amplitudes equal to zero, forcing all such amplitudes to be zero. In the present case of infinite-dimensional representations of the noncompact de Sitter symmetry group, Ward identities do not close on a finite set of equations but produce recurrence relations. In principle, they admit solutions with
$\langle{\scriptstyle (000)(000)(000)\big|0 }\rangle\neq 0$, indicating the existence of a linear combination (of an infinite number of simple products of one-particle states) which includes ${\scriptstyle (000)(000)(000)}$
and is invariant under the symmetry transformations. Indeed, such states were identified in \cite{rep4}, but were shown to be non-normalizable and therefore unphysical. The vanishing of $\langle{\scriptstyle (000)(000)(000)\big|0 }\rangle$ 
shows that the results of \cite{Taylor:2025deepIR}, based on the formalism developed in \cite{Taylor:2024vdc},  agree with the group-theoretic anal{\nolinebreak}ysis of \cite{rep4}.
Note that the derivation of (\ref{phider}) never used the particular combination of the Ferrers functions (\ref{lwap}) that defines the Bunch-Davies vacuum and leads to the vanishing of ``all-out'' amplitudes. Therefore, Ward identities also hold in unphysical $\alpha$-vacua \cite{Mottola:1984ar,Allen:1985ux}.

Ward identities can be also used for deriving ``crossing'' relations, with some particles crossing from the initial to final states and vice versa. As an example, starting from
\be \langle 0|\,[\,Q_{\gamma},a_{110}a_{000}a_{000}\, Ua^\dagger_{000}\,]\,|0\rangle=0\ ,\ee
we obtain
\be\big\langle{\scriptstyle(110)(000)(000)\big|(110) \big\rangle}=2\,\big\langle{\scriptstyle(110)(000)(1{-}10)\big|(000)  \big\rangle} +  \big\langle{\scriptstyle(000)(000)(000)\big|(000) \big\rangle}  .  \ee
Another similar example is
\begin{equation}
\begin{aligned}\big\langle{\scriptstyle(110)(110)(1{-}10)\big|(110) \big\rangle}    &  -2 \,\big\langle{\scriptstyle (110)(1{-}10)(000)\big|(000) \big\rangle}        \\[1mm]
 = ~&{\textstyle  \sqrt{\frac{4}{3}\left(\frac{\mu^2 + \frac{25}{4}}{\mu^2+\frac{9}{4}}\right)} }\Big( \big\langle{\scriptstyle (110)(1{-}10)(200)\big|(000) \big\rangle}       + \big\langle{\scriptstyle (110)(110)(2{-}20)\big|(000) \big\rangle}  \Big),
\end{aligned}
\end{equation}
which follows from
\be \langle 0|\,[\,Q_{\gamma},a_{110}a_{110}a_{1{-}10}\, Ua^\dagger_{000}\,]\,|0\rangle=0\ .\ee

\subsection{Flat limit}
In the limit of short wavelengths and high frequencies, scattering processes probe small invariant spacetime intervals. Expanding around an observer at the North pole, $t\approx\chi\approx\theta\approx 0$, and
\be (X^0, X^1,X^2,X^3,X^4)~\approx~ (t, \,1, \,\theta,\,\chi\cos\varphi,\,\chi\sin\varphi).\ \ee
In this neighborhood, the de Sitter metric is approximately flat,
\be ds^2\approx -dt^2+d\theta^2+\chi^2d\varphi^2, \ee
with the hyperplane tangent to $S^3$ parametrized by cylindrical coordinates $z,\rho,\varphi$ identified as
\be \theta=z\ ,  \quad\chi=\rho\ , \quad \varphi=\varphi\ \label{cyl1}.\ee

The wavefunctions describing short wavelengths are related to the large $k$ limit of hyperspherical harmonics. In this limit, the In\"on\"u-Wigner contraction \cite{ew} identifies $k$ as the magnitude of the spatial momentum $\bm{p}$, $k=|\bm{p}|$. We want to show that as $k\to\infty$, de Sitter Ward identities describe the Poincar\'e symmetry of the flat limit. To that end, we need to change the basis of asymptotic states from the angular momentum to the standard plane wave basis. This may be done for arbitrary spin but here we focus on the scalar principal series. Plane waves can be expressed in terms of cylindrical waves using the Jacobi-Anger expansion:
\begin{equation}
e^{i \bm{p}\cdot \bm{r}} =  e^{ip_z z}\sum_{m = -\infty}^{\infty}i^m e^{- i m\alpha}J_m(\rho\, p_{\rho})e^{i m\varphi} = e^{ip_z z}\sum_{m = -\infty}^{\infty}i^M e^{- i m\alpha}J_M(\rho\, p_{\rho})e^{i m\varphi}, \label{cylwaves}
\end{equation}
where $J_m $ are the Bessel functions of the first kind with order $m$ and the momentum components
\be \bm{p}~=~\big( p_x=p_\rho\cos\alpha,\, p_y=p_\rho\sin\alpha,\, p_z\big).\  \ee 
It should be noted that 
\begin{equation}
\lim_{M\rightarrow\infty}J_M(\rho \, p_{\rho}) \sim \frac{1}{\sqrt{2\pi M}}\left(\frac{e\,\rho \, p_{\rho}}{2M}\right)^M,
\end{equation}
so terms in (\ref{cylwaves}) with large $M = |m|$ are suppressed.

Our angular momentum basis consists of the hyperspherical harmonics written in (\ref{harms}). We are interested in the ``relativistic'' (high-energy) limit
\be E\equiv k\gg \mu\ , ~~~~\frac{L}{k}~\,\makebox{and}~\frac{M}{k}~\makebox{fixed},\ee
in the region of small $\chi$. In this limit, $d\gg1$ and we can use the asymptotic formula \cite{Magnus:Formulas}
\begin{equation}
\lim_{d \rightarrow \infty} d^{-M}P_d^{(M, L)}(\cos{2\rho}) = (d\rho)^{-M}J_{M}(2d\rho).
\end{equation}
Furthermore, the normalization constants $N_{kl m}$ become
\begin{equation}
N_{k l m} \approx \frac{i^M}{i^m}\sqrt{\frac{k}{2\pi^2}} \ ,
\end{equation} hence
\begin{equation}
Y_{kl m}(\Omega) \approx  \frac{i^M}{i^m}\sqrt{\frac{k}{2\pi^2}}\,
J_{M}(2d\chi)\, e^{il \theta}\, e^{i m \varphi}.
\label{hexp}
\end{equation}
With the cylindrical coordinates identified as in (\ref{cyl1}), the harmonics acquire the form of the cylindrical wave in (\ref{cylwaves}) once we identify the energy-momentum components as
\be k=E\ , \qquad l=p_z\  ,\qquad 2d \approx p_{\rho}=\sqrt{k^2-l^2}.\label{momids}\ee
The plane wave can now be written as
\begin{equation}e^{i\bm{p}\cdot \bm{r}}
   = \pi \sqrt{\frac{2}{k}} \sum_{m = -\infty}^{\infty}i^me^{-im\alpha}Y_{klm}(\Omega),
\label{pwave}\end{equation}
and we can relate the momentum bases in angular and linear coordinates via the Fourier transform
\begin{equation}|\,\bm{p}\,\rangle
   = \pi \sqrt{\frac{2}{k}} \sum_{m = -\infty}^{\infty}i^{m}e^{-im\alpha}\,|k\,l\,m\,\rangle ,
\label{pst}\end{equation}
with the momentum angle $\alpha$  conjugate to $m$. Hence $m$ is the angular momentum in the $z$ direction. The observer's neighborhood, i.e.\ the region  of small $\rho$ (and $z$), is probed by the waves with small $|m|/k\sim\rho$. Therefore in the flat limit, we focus on  $|m|=M\ll k$.

The ``high-energy'' scattering amplitudes are defined as
\begin{align}
\big\langle\bm{p_{r+1}} {\dots}\bm{p_{N}}& \big|\, \bm{p_{1}}{\dots}\bm{p_r}\big\rangle= \label{ftr}\\ &\Big(\prod_{a=1}^{r}\pi \sqrt{\frac{2}{k_a}}  \sum_{m_a= -\infty}^{\infty} i^{m_a}e^{-im_a\alpha_a}\Big)   \Big(\prod_{b=r+1}^{N}\pi \sqrt{\frac{2}{k_b}} \sum_{m_b= -\infty}^{\infty}i^{-m_b}e^{im_b\alpha_b}\Big)\nonumber \\[2mm]
&~~~\times\big\langle(k_{r+1},l_{r+1},m_{r+1})\dots (k_N,l_N,m_N)\big|(k_1,l_1,m_1)\dots (k_r,l_r,m_r)\big\rangle\ .
\nonumber\end{align}

We start from Ward identities in the angular basis and apply the Fourier transforms as in (\ref{ftr}). The identities associated to the isospin charges $Q_L$ and $Q_{L'}$ read
\be \Big[\sum_{a=1}^r(l_a\pm m_a)-\sum_{b=r}^N(l_b\pm m_b)\Big]\big\langle(\scriptstyle k_{r+1},\scriptstyle l_{r+1},m_{r+1})\dots (k_N,l_N,m_N)\big|(k_1,l_1,m_1)\dots (k_r,l_r,m_r)\big\rangle\displaystyle =0\ .\ee
After Fourier transforming and identifying $l=p_z$, we obtain
\be\Big[\sum_{a=1}^r\big(p_{az}\mp i\frac{\partial}{\partial\alpha_a}\big)-\sum_{b=r}^N  \big(p_{bz}\pm i\frac{\partial}{\partial\alpha_b}\big)   \Big]\big\langle\bm{p_{r+1}} {\dots}\bm{p_{N}}\big|\, \bm{p_{1}}{\dots}\bm{p_r}\big\rangle=0\ .\ee
This is the conservation law for the momentum and angular momentum components in the $z$ direction.

Other Ward identities involve quantum numbers $k,l,m$ shifted by $\pm 1$. For large $k$ and $l$, finite differences can be replaced by derivatives. For example, in the Ward identity associated with $Q_{\gamma}$, we encounter
\begin{equation}\big\langle{\dots} \big|\dots (k\pm1,l+1,m){\dots} \big\rangle\approx \Big(1\pm\frac{\partial}{\partial k}  +\frac{\partial}{\partial l}  \Big)
\big\langle{\dots} \big|\dots (k,l,m){\dots} \big\rangle.\label{der1}
  \end{equation}
After Fourier transforming and with the identifications (\ref{momids}), we find
\be\frac{\partial}{\partial k}=\frac{\partial}{\partial E}+\frac{E}{p_{\rho}}\frac{\partial}{\partial p_\rho}\ ,
\quad  \frac{\partial}{\partial l}=\frac{\partial}{\partial p_z}-\frac{p_z}{p_{\rho}}\frac{\partial}{\partial p_\rho}.      \label{der2}           \ee
Note that in the linear momentum space, shifting $m\to m\pm 1$ amounts to multiplication by $\mp ie^{\pm i\alpha}$, and multiplicative factors of $m$ become derivatives ${-}i\tfrac{\partial}{\partial \alpha}$. 

As an example, consider the Ward identities associated with $Q_{\pm\gamma}$. Recall that  $Q_{\gamma}$ raises $l$ by 1, while raising or lowering $k$ by 1, with the coefficients $A_{\gamma}$ and $D_{\gamma}$, respectively, while $Q_{-\gamma}$ lowers $l$ by 1. In the relativistic limit,
\be A_{\pm\gamma}(k,l,m)\approx -\frac{i}{2}(k\pm l)=-\frac{i}{2}(E\pm p_z)\ ,~D_{\pm\gamma}(k,l,m)\approx -\frac{i}{2}(k\mp l)=-\frac{i}{2}(E\mp p_z).\ \ee
After taking the limits of coefficients and using (\ref{der1}) and (\ref{der2}), individual terms in the  $Q_{\pm\gamma}$ Ward identities aquire the form
\begin{align}A_{\pm\gamma}(k,l,m)&\big\langle{\dots} \big|\dots (k+1,l\pm 1,m){\dots} \big\rangle +D_{\pm\gamma}(k,l,m)\big\langle{\dots} \big|\dots (k-1,l\pm 1,m){\dots} \big\rangle\nonumber\\[1mm] &~~~=~-\frac{i}{2}\Big[E\pm \big(p_z\frac{\p}{\p E}+E\frac{\p}{\p p_z}\big)\Big]\big\langle{\dots} \big|\dots (k,l,m){\dots} \big\rangle\ .
  \end{align}
These Ward identities impose energy conservation and invariance under Lorentz boosts in the $z$ direction. In a similar way, one can show that the remaining de Sitter generators impose full Poincar\'e symmetry of the high-energy amplitudes (\ref{ftr}).

\section{Conclusions}
In this work, we studied the Hilbert space of particles of the standard model and gravitons in global de Sitter spacetime. Massive particles  belong to the principal and complementary series representations of de Sitter symmetry group while gauge bosons and gravitons belong to the discrete family. Although the properties of these unitary representations have already been discussed from the mathematical perspective by Dixmier in 1961, not much attention has been paid to their implications for the physical processes occurring in de Sitter spacetime, particularly the role of de Sitter isometry in the scattering processes.

{}From the symmetry point of view, the most natural
set of coordinates parametrizing de Sitter's $S^3$ directions are not the spherical coordinates, but the Hopf angles (toroidal coordinates). We showed that when the scalar wavefunctions, i.e.\ the solutions of Klein-Gordon equation, are expressed in terms of these coordinates, they transform in a simple way under de Sitter symmetry transformations. In this way, we were able to give a ``physicist's'' derivation of Dixmier's results for the principal series. We also discussed the representations associated to spin $\frac{1}{2}$ fermions, gauge bosons, and gravitons.

In Minkowski spacetime, massless particles, including photons, behave in a rather peculiar way. The scattering amplitudes involving their zero frequency modes contain infrared divergences, a troublesome feature that marred the S-matrix theory from its inception. There is no such problem in de Sitter spacetime. The UIRs describing gauge bosons and gravitons have no zero frequency modes. Even in scalar field theory, the zero frequency modes decouple in the zero mass limit.

The main result of this paper is the Ward identities for de Sitter scattering amplitudes. They were obtained by using the transformation properties of quantum field operators and one-particle states
and are completely universal, because the only assumption used in the derivation was the existence of a de Sitter invariant time evolution operator.
They have the form of recurrence relations. They are as restrictive as Poincar\'e Ward identities in flat spacetime, with the angular momentum conservation on $S^3$ similar to the linear momentum conservation in Minkowski space. There are some processes, however, that are forbidden in flat spacetime but are allowed in de Sitter. For example, there are no kinematic restrictions on particle decays so de Sitter particles are generically unstable in any interacting theory, unless their stability is protected by internal symmetries. Furthermore, massless or even massive particles can in principle be spontaneously produced from the vacuum. However, as found in \cite{Taylor:2025deepIR}, the corresponding amplitudes are zero. The absence of particle production from the vacuum can  be explained using Ward identities and general group-theoretic arguments, and therefore is a consequence of de Sitter symmetry.

We discussed the ``flat'' limit of Ward identities, for the processes with short wavelengths and/or high frequencies. In this limit, the toroidal coordinates are identified with cylindrical coordinates and the Ward identities reflect the Poincar\'e symmetry of short-distance physics. As a result, the decays of particles with Compton wavelengths shorter than  de Sitter radius, as well as other processes disallowed by Poincar\'e symmetry, are  strongly (exponentially) suppressed.

There are several directions for future work on the symmetries of de Sitter amplitudes. In our opinion, the most important task is the construction of  kinematic invariants, analogous to Mandelstam's variables, that would automatically ensure de Sitter symmetry of the scattering amplitudes. It is also important to understand the relation between the amplitudes evaluated in global de Sitter spacetime and the cosmological correlators \cite{Chen:2009zp, Arkani-Hamed:2015bza}  studied in the Poincar\'e patch.

\section*{Acknowledgments}
AL is deeply grateful to Sophie (cat) for her steady contributions to morale. This work was supported in part by NSF PHY-2209903, the Simons Collaboration on Ce{\nolinebreak}lestial Holography, and
Polish National Agency for Academic Exchange under the NAWA Chair programme.
It was also supported by the MAESTRO grant no.\ 2024/54/A/ST2/00009
funded by the National Science Centre, Poland. 
Any opinions, findings, and conclusions or
recommendations expressed in this material are those of the authors and do not necessarily
reflect the views of the National Science Foundation.
\appendix
\section{Killing vectors in global coordinates}
We collect here the Killing vectors that generate global de Sitter isometries. They follow directly from the embedding space definition $K_{AB} = X_A \partial_B - X_B \partial_A$ and the coordinate definitions (\ref{eq:coordinates}).
\begin{equation}
\begin{aligned}
K_{01} & = \cos{t}\cos{\chi}\cos{\theta}\;\partial_t - \sin{t} \sin{\chi} \cos{\theta} \;\partial_{\chi} - \sin{t}\sec{\chi}\sin{\theta}\;\partial_{\theta},\\
K_{02} & = \cos{t}\cos{\chi}\sin{\theta} \;\partial_t - \sin{t} \sin{\chi} \sin{\theta} \;\partial_{\chi} + \sin{t}\sec{\chi}\cos{\theta}\;\partial_{\theta}, \\
K_{03} & = \cos{t}\sin{\chi}\cos{\varphi} \;\partial_t + \sin{t} \cos{\chi} \cos{\varphi} \;\partial_{\chi} - \sin{t}\csc{\chi}\sin{\varphi}\;\partial_{\varphi}, \\
K_{04} & = \cos{t}\sin{\chi}\sin{\varphi} \;\partial_t + \sin{t} \cos{\chi} \sin{\varphi} \;\partial_{\chi} + \sin{t}\csc{\chi}\cos{\varphi}\;\partial_{\varphi},\\ 
K_{12} & = \partial_{\theta},\\
K_{13} & = \cos{\theta}\cos{\varphi}\;\partial_{\chi} + \tan{\chi}\sin{\theta}\cos{\varphi}\;\partial_{\theta} - \cot{\chi}\cos{\theta}\sin{\varphi}\;\partial_{\varphi},\\
K_{14} & = \cos{\theta}\sin{\varphi}\;\partial_{\chi} + \tan{\chi}\sin{\theta}\sin{\varphi}\;\partial_{\theta} + \cot{\chi}\cos{\theta}\cos{\varphi}\;\partial_{\varphi},\\
K_{23} & = \sin{\theta}\cos{\varphi}\;\partial_{\chi} - \tan{\chi}\cos{\theta}\cos{\varphi}\;\partial_{\theta} - \cot{\chi}\sin{\theta}\sin{\varphi}\;\partial_{\varphi},\\
K_{24} & = \sin{\theta}\sin{\varphi}\;\partial_{\chi} - \tan{\chi}\cos{\theta}\sin{\varphi}\;\partial_{\theta} + \cot{\chi}\sin{\theta}\cos{\varphi}\;\partial_{\varphi},\\
K_{34} & = \partial_{\varphi}.
\end{aligned}
\end{equation}

\section{Ferrers functions and Jacobi polynomials}

The computations leading to (\ref{phider}) rely on several identities for the Ferrers functions $P^{\alpha}_{\beta}(w), Q^{\alpha}_{\beta}(w)$ and Jacobi polynomials $P^{(M, L)}_d(u)$. The $\theta$- and $\varphi$-dependence is simple, so only the $\chi$- and $t$-dependent parts of our mode functions $\phi^{(+)}_{k l m}(t, \Omega)$ require nontrivial simplifications. We record here the identities used repeatedly in the calculations. Throughout we set
\begin{equation}
    w = \sin{t} \qquad u = \cos{2\chi}.
\end{equation}

\subsection{Ferrers functions}

For $x \in (-1, 1)$, the associated Legendre equation
\begin{equation}
    (1-x^2)\frac{d^2y}{dx^2} - 2x\frac{dy}{dx}+ \left(\beta(\beta+1) - \frac{\alpha^2}{1-x^2}\right)y = 0
\end{equation}
has linearly independent solutions denoted by $P^{\alpha}_{\beta}(x)$ and $Q^{\alpha}_{\beta}(x)$, which are known as the associated Legendre functions on the cut (of the first and second kind, respectively), or simply Ferrers functions. Writing $x = \sin{t}$ and $\gamma = (\tfrac12 + \beta)(\tfrac{\pi}2 - t) + (\tfrac12 + \alpha)\tfrac{\pi}{2}$, we can express $P^{\alpha}_{\beta}(x)$ and $Q^{\alpha}_{\beta}(x)$ in terms of the hypergeometric function $\,{}_2F_1(a,b;c;z)$ as
\begin{equation}
\begin{aligned}
P^{\alpha}_{\beta}(\sin{t})
= {} & \frac{i}{\sqrt{2\pi\,\cos{t}}}
      \frac{\Gamma(1+\alpha+\beta)}{\Gamma\!\left(\tfrac{3}{2}+\beta\right)}
      \Bigg[
        e^{-i\gamma}\,
        {}_2F_1\!\left(
          \tfrac12+\alpha,\,
          \tfrac12-\alpha;\,
          \tfrac32+\beta;\,
          \frac{1}{2\cos{t}}
          e^{it}
        \right)
\\[6pt]
&\qquad
      -\, e^{i\gamma}\,
        {}_2F_1\!\left(
          \tfrac12+\alpha,\,
          \tfrac12-\alpha;\,
          \tfrac32+\beta;\,
          \frac{1}{2\cos{t}}
          e^{\,{-}it}
        \right)
      \Bigg],
\end{aligned}
\end{equation}

\begin{equation}
\begin{aligned}
Q^{\alpha}_{\beta}(\sin{t})
= {} & \frac{1}{2}\frac{\pi}{\sqrt{2\,\cos{t}}}
      \frac{\Gamma(1+\alpha+\beta)}{\Gamma\!\left(\tfrac{3}{2}+\beta\right)}
      \Bigg[
        e^{-i\gamma}\,
        {}_2F_1\!\left(
          \tfrac12+\alpha,\,
          \tfrac12-\alpha;\,
          \tfrac32+\beta;\,
          \frac{1}{2\cos{t}}
          e^{it}
        \right)
\\[6pt]
&\qquad
      +\, e^{i\gamma}\,
        {}_2F_1\!\left(
          \tfrac12+\alpha,\,
          \tfrac12-\alpha;\,
          \tfrac32+\beta;\,
          \frac{1}{2\cos{t}}
          e^{-it}
        \right)
      \Bigg].
\end{aligned}
\end{equation}
These expansions are most natural for recognizing the linear combination in (\ref{lwap}) as a positive frequency mode in the large momentum limit.


The following identities account for all $k \rightarrow k \pm 1$ shifts appearing in the action of the generators $X_{\pm \gamma}, X_{\pm \delta}$ (with identical formulas for $Q^{\alpha}_{\beta}(w)$) \cite{Magnus:Formulas}:
\begin{equation}
\begin{aligned}
w\; P^{\alpha}_{\beta}(w) & = \frac{1}{2\beta + 1}\left((\beta-\alpha + 1)P^{\alpha}_{\beta+1}(w) + (\beta + \alpha)P^{\alpha}_{\beta-1}(w)\right)\\
(1-w^2)\;\partial_w \;P^{\alpha}_{\beta}(w) & = \frac{1}{2\beta+1}\left(-\beta(\beta - \alpha +1)P^{\alpha}_{\beta + 1}(w) + (\beta + 1)(\beta+\alpha)P^{\alpha}_{\beta - 1}(w)\right).\\
\end{aligned}
\end{equation}
\subsection{Jacobi polynomials}
\label{subsec:jacobi}
The Jacobi polynomials $P_d^{(M, L)}(u)$ are the set of orthogonal polynomials on the interval $(-1, 1)$ with weight function $W(u) = (1-u)^{M}(1+u)^{L}$, and satisfy the differential equation
\begin{equation}
(1-u^2)\,\frac{d^2y}{du^2} + \left(M-L+(M+L-2)\,u\right)\frac{dy}{du} + (d+1)(M+L+d)\,y = 0.
\end{equation}
They are often expressed in terms of the hypergeometric function $\,{}_2F_1(a,b;c;z)$ as
\begin{equation}
P_d^{(M,L)}(u) = {d + M \choose M} \,{}_2F_1\left(-d, d+L+M+1; M+1; \frac{1-u}{2}\right). 
\end{equation}
Differentiating $\phi_{k l m}(t, \Omega)$ with respect to $\chi$ produces a term proportional to 
\begin{equation}
\begin{aligned}
\partial_{\chi}(\cos^L{\chi}\sin^M{\chi}P_d^{(M, L)}(u)) = & \cos^L{\chi}\sin^M{\chi}(-L\tan{\chi}+M\cot{\chi})P_d^{(M, L)}(u)\\
&-2(d+M+L+1)\cos^{L+1}{\chi}\sin^{M+1}{\chi}P_{d-1}^{(M+1, L+1)}(u).
\end{aligned}
\end{equation}
To rewrite the full expression for the action of a generator on $\phi_{k l m}(t, \Omega)$ in terms of  
\begin{equation}
\cos^{L'}{\chi}\sin^{M'}{\chi}P_{d'}^{(M', L')}(u)
\end{equation}
with shifted parameters $M', L', d' = \frac{1}{2}(k' - M' - L')$, we use identities of the form \cite{Magnus:Formulas}
\begin{equation}
(d+L)P_d^{(M+1, L-1)}(u) = LP_d^{(M, L)}(u) + \frac{1}{2}(d+M+L+1)(1+u)P_{d-1}^{(M+1, L+1)}(u).
\end{equation}
Most of the expressions involving Jacobi polynomials were simplified in Mathematica using identities of this form; we include this identity here as a representative example since it was the only case not recognized automatically by Mathematica (when $M$ and $L$ shift in opposite directions).

\end{document}